\documentclass[twocolumn]{aastex63}

\usepackage{graphicx}
\usepackage{amsmath}
\usepackage{amssymb}
\usepackage{natbib}
\usepackage{color}
\usepackage{enumitem}
\usepackage[T1]{fontenc}

\usepackage{txfonts}
\let\tablenum\relax 
\usepackage{siunitx}

\input{newcommands.sty}

\newcommand{\corrected}[1]{{#1}}
\newcommand{\correctedold}[1]{{#1}}

\received{April 21, 2020}
\revised{July  8 and July 20, 2020}
\accepted{\today}
\submitjournal{ApJ}

\shorttitle{Diffusion limited planetesimal formation}
\shortauthors{Klahr and Schreiber}

\begin{document}

\title{Turbulence sets the length scale for planetesimal formation:\\
Local 2D simulations of streaming instability and planetesimal formation}

\correspondingauthor{Hubert Klahr}
\email{klahr@mpia.de}

\author[0000-0002-8227-5467]{Hubert Klahr}
\author{Andreas Schreiber}
\affil{Max Planck Institut f\"ur Astronomie, K\"onigstuhl 17, 69117, Heidelberg, Germany}



\begin{abstract}
The trans-Neptunian object 2014 MU69, named Arrokoth, 
is the most recent evidence that planetesimals did not form by successive collisions of smaller objects, but by the direct gravitational collapse of a pebble cloud. But what process sets the physical scales on which this collapse may occur? Star formation has the Jeans mass, that is when gravity is stronger than thermal pressure, helping us to understand the mass of our sun. But what controls mass and size in the case of planetesimal formation? 
Both asteroids and Kuiper belt objects show a kink in their size distribution at 100 km.
Here we derive a gravitational collapse criterion for a pebble cloud to fragment to planetesimals, showing that a critical mass is needed for the clump to overcome turbulent diffusion. 
We successfully tested the validity of this criterion in direct numerical simulations of planetesimal formation triggered by the streaming instability.
Our result can therefore explain the sizes for planetesimals found forming in streaming instability simulations in the literature, while not addressing the detailed size distribution.
We find that the observed characteristic diameters of $\sim$ 100 km corresponds to the critical mass of a pebble cloud set by the strength of turbulent diffusion stemming from streaming instability for a wide region of a solar nebula model from 2 - 60 au, with a tendency to allow for smaller objects at distances beyond and at late times, when the nebula gas gets depleted. 
\end{abstract}



\section{Introduction} \label{sec:intro}
One of the classical ideas in planet formation is to form planetesimals in a direct gravitational collapse of pebble clouds \citep{Safronov1969,GoldreichWard1973}. Most planetesimals were incorporated into planetary bodies, yet the asteroids, Kuiper Belt objects (KBOs) and comets are believed to be leftovers from the initial plethora of planetesimals. A study of these minor bodies in the solar system is therefor a key to understand planetesimal and ultimately planet formation. One problem is to subtract 4.5 Billion years of collisional evolution of planetesimals from the size distribution found today.

Recent observational work \citep{Delbo2017} 
identified a group of asteroids that clearly did not originate as collisional fragments of larger ones. 
Members of this asteroid group are all larger than 35 km, with a most likely diameter of $\sim$100 km, \correctedold{confirming the previous assumption that Planetesimals are born big \citep{Morbidelli2009} and that the 100 km bump in the size-frequency-distribution (SFD) is primordial.}  Also the impact size distribution on the surfaces of Pluto and its largest moon Charon, as recently determined in the New Horizon mission \citep{Singer2019}, finds a strong deficiency of Kuiper belt objects \correctedold{smaller than $1-2$ km in size, currently interpreted as an effect of collisional grinding.}
The bi-lobed structure of \correctedold{Arrokoth} \citep{Stern2019}, with both parts of quite similar material and the general high fraction of binaries among Kuiper belt objects is further support for a gravitational collapse scenario for planetesimal formation. \correctedold{As \citet{2019NatAs.tmp..415N} have shown, binaries formed in the streaming instability scenario, have a specific distribution of inclinations, which turns out to be a good match to observed binaries.}

The gravitational collapse of pebble clouds, i.e.\ accumulations of about cm sized solid material, in the solar nebula is indeed a rapid and efficient route to form planetesimals \citep{JohansenKlahrHenning2006,2007Natur.448.1022J}. 
The definition of pebbles in planet formation is thus not simply \correctedold{based on their} size, but rather given via their aerodynamic properties expressed as a Stokes number $\St = \tau_{\rm f} \Omega$, i.e. the product of aerodynamic "friction" or "stopping" time $\tau_{\rm f}$
and the local Keplerian frequency $\Omega$. 

The Stokes number is a central parameter in planetesimal formation, because it not only determines the radial drift, but also the vertical sedimentation speed $v_z = \St \Omega z$ at height $z$ above the midplane, as well as the efficiency with which particles couple to the gas turbulence in the disk, which most likely has a correlation time at large scales on the order of $1/\Omega$. Also the strength of instabilities related to the particle feedback onto the gas relates to the Stokes number \citep{2018MNRAS.477.5011S}, as for instance the streaming instability \citep{YoudinGoodman2005}.

Our numerical simulations \correctedold{\citep{Birnstiel2012} of} particle growth and disk evolution have shown that $\St = 0.01 - 0.1$ should be the typical value for the largest expected grains, which then dominate the radial influx of dust and ice grains. 

The goal of this paper is to put an absolute length- or mass-scale onto the planetesimal formation process \correctedold{. Because asteroids and KBOs show both a bump in the SFD at 100 km, this scale should not strongly} depend on the distance from the central star, but only on the mass of that star and the properties of pebbles in dimension free Stokes numbers and the strength of turbulent diffusion also in a dimension free version like the $\alpha$ description. 

If one compares planetesimal formation to star formation, then this would be like the derivation of a Jeans-mass for planetesimals. The Jeans-mass for stars also reflects local properties of the gas, like the local gas temperature and density, helping us to understand why certain stellar masses are more likely than others. Yet stars come at both higher and lower than 1 Jeans-mass. As with planetesimals there is an initial mass function for stars, generated by physical processes still under debate until today \citep{2014prpl.conf...53O}. Is it competitive accretion after fragmentation of the initial unstable cloud core or the power-spectrum of the turbulence that dictates the shape of the initial mass function, or some combination or something else? Already the determination of the initial stellar mass function is a problem towards the higher masses, as those stars don't live for too long and it may be difficult to identify multiple systems. Same is true for planetesimals, as today one predominantly only find the left overs after incorporating most material into planets and having had the planetesimals undergo a 4.5 billion year lasting collisional evolution. Numerical simulations of star and planetesimal formation both find power laws and the concept of either Jeans-mass or the here derived concept of a diffusion-mass puts a scaling on the numerical experiments.

In the following section we summarise our current understanding of planetesimal formation via gravitational instability via turbulent clustering, via trapping in zonal flows as well as in the bump free streaming instability scenario. In Section 3 we derive our Jeans like stability criterion based on comparing turbulent diffusion and gravitational contraction timescales. We test this stability criterion successfully in Section 4 by a parameter study on planetesimal formation in a streaming instability scenario. 
We explore the resulting planetesimal sizes for turbulence values in the solar nebula in Section 5 and find indeed 100 km as a realistic value for a \correctedold{equivalent diameter} of planetesimal \correctedold{forming pebble clouds}  all over the solar nebula. We briefly discuss and interpret our results in Section 6 and shift all technical details to the appendices of this paper.
\begin{figure*}
\begin{center}
\includegraphics[width=0.80\textwidth]{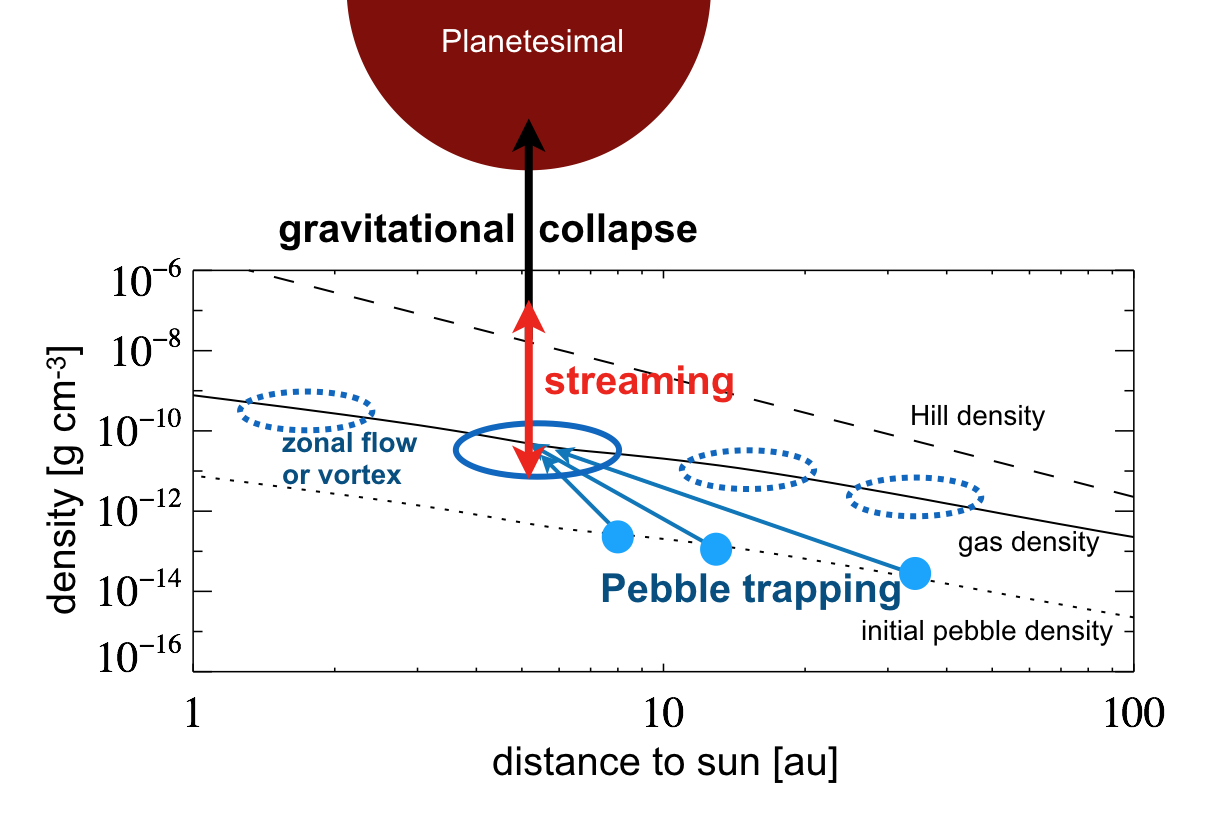}
\caption{Drift-Trap-Stream-Collapse (schematic representation of densities): In our paradigm the sequence of planetesimal formation starts with particles growing to pebble size, they sediment to the midplane and drift towards the star to get \correctedold{temporarily} trapped in a zonal flow or vortex \citep{Lenz_2019}. Here the density of pebbles in the midplane is eventually sufficient to trigger the streaming instability. This instability concentrates and diffuses pebbles like wise, leading to clumps that reach the Hill density at which tidal forces from the star can no longer shear the clumps away.
If now also the turbulent diffusion is weak enough to let the pebble cloud collapse, then planetesimal formation will occur.
\label{Fig:drift-trap}}
\end{center}
\end{figure*}

\section{Planetesimal formation via self gravity}

It is well understood that if local particle traps in a turbulent disk stop or reduce the radial drift sufficiently, then significant local overdensities can occur which will be able collapse under their own gravitational attraction \citep{JohansenKlahrHenning2006}, a process we coin gravito turbulent planetesimal formation, as it resembles similar processes in star formation.
In \citep{2007Natur.448.1022J}, simulating a fully turbulent disk by means of the magneto rotational instability creating zonal flows, it was shown that trapping leads to 
a rapid formation of planetesimals, with and without including particle feed back onto the gas. For simulations without the background turbulence and trap formation, the typical solar metalicity was insufficient, as Kelvin Helmholtz and Streaming Instability prevents particles from the necessary sedimentation \citep{2007Natur.448.1022J, Gerbig2020}. 

\cite{Johansen2014} showed that only a dust enrichment by a factor of $2-3$ 
above the average abundance of solids in the solar nebula, will lead to gravitational collapse in the presence of streaming instability without background turbulence respectively the formation of particle traps in the solar nebula \citep{2018haex.bookE.138K}.
In that case the local dust enrichment could be produced by photo-evaporation of the disk gas in the later stages of disk evolution \citep{Carrera2017}. 

If the the local dust enrichment was the result of different processes, e.g.\ dust trapping in vortices or zonal flows \citep{JohansenKlahrHenning2006,2007Natur.448.1022J,Dittrich2012,Raettig2014} then it would start much earlier in the evolution of the solar nebula \citep{Lenz_2019}, but it is currently not clear if this enrichment in pebbles would first have to trigger a streaming instability, or whether the enrichment would directly fragment once it overcomes diffusion by Kelvin Helmholtz and Streaming Instability \citep{Gerbig2020}. 

It should also be mentioned that if there was no radial pressure gradient in the disk, no streaming instability would occur and one would directly go into gravitational collapse in a laminar disk even for lower dust enrichment at a so far undetermined level \citep{Abod2018}. Streaming instability is no precondition for planetesimal formation, but rather can be the controlling agent of efficiency and as we show in this paper, by defining a threshold mass for planetesimal formation.

Numerical simulations in which a relatively large volume, possibly larger than a trapping region might be, was globally enriched in particles, have lead to the derivation of various characteristic power laws in the size distribution of planetesimals \citep{Johansen2015,Simon2016,Simon2017}. In the highest resolution cases a deficiency of small planetesimals seems to appear, which could correspond to the critical mass for collapse as derived in our paper, yet unfortunately the turbulence diffusivity was not measured in those simulations, which would be the controlling parameter as we will show.

The power laws of the size distribution as found in the literature are rather shallow \correctedold{\citep{Johansen2015, Simon2016, Schaefer2017}.
Most recently \citet{Abod2018} report a mass distribution $dN/dM_{\rm P} \propto M_{\rm P}^{-p}$ with a value of $p \approx 1.6$ for a range of pressure gradients. If one uses an exponentially truncated powerlaw, even $p \approx 1.3$ makes a good fit for the low mass end.}
Thus the mass dominating planetesimal size in these simulations is typically the largest object, i.e.\ several hundred kilo-meters, reflecting the total mass of pebbles that was initially put into these simulations. As a result these simulations explore rather the largest possible planetesimals for a favourable scenario of global enrichment in pebbles. 

Interestingly also in the cascade model for planetesimal formation by turbulent clustering, in its latest version described in \citet{HartlepCuzzi2020}, a locally enhanced amount of pebbles and a reduced headwind as in a zonal flow is needed, to form a sufficient number of planetesimals within the lifetime of the solar nebula. In this model it is not trapping of pebbles in pressure maxima or concentration as effect of the streaming instability, but a random concentration event in the gas turbulence of the disk. Random concentrations that exceed the local Hill density to form a gravitationally bound planetesimal are rare, and it appears to need a fine tuning of Stokes number, turbulence strength, and more importantly a gas density enhancement factor, a solids enhancement factor, and a scale factor for the headwind parameter to get the desired amount and the sizes of planetesimals within 2 Million years \citep{HartlepCuzzi2020}. 

\correctedold{The three scenarios discussed above, trapping, streaming and clustering are not mutually exclusive as they all involve a local enhancement of material, involving particle traps like zonal flows and vortices. Thus they can all be parameterized in planetesimal formation rate as a function of local influx of pebbles as we did in \citep{Lenz_2019, gerbig2019, Lenz2020}, yet at different conversion efficiencies from pebbles to planetesimals.

It is now our claim that in all these three scenarios, the final collapse of a cloud is regulated by turbulent diffusion on the scales of self-gravitating pebble clouds. 
In Fig.\ \ref{Fig:drift-trap} we sketch the general scenario of pebble trapping and converting them locally into planetesimals by gravitational collapse, this general picture should hold for any gravity assisted planetesimal formation scenario. 

Building on that paradigm we ask now for the minimal mass in pebbles that has to be locally accumulated to trigger gravitational collapse. This is different from the ansatz in \cite{drazkowska2016} and \cite{schoonenberg2018} as they ask for the critical density or dust to gas ratio in the midplane to trigger streaming instability and thus planetesimal formation, whereas one should ask to concentrate pebbles to Hill density and a critical mass. Yet it is a good criterion to ask for dust-to-gas of one in the miplane, as a critical stage, because one has to overcome this critical limit on the way to Hill density anyway and once we are past dust-to-gas of 1, the effect of turbulence gets diminished anyway \citep{2007Natur.448.1022J}}

\section{A Jeans like length scale criteria for gravitational collapse of pebble clouds}

We can formulate a collapse criterion that requires a critical clump size $r = \lcrit$ (i.e. the radius of pebble cloud) to be large enough to contract against any sort of underlying turbulent diffusion $D$.
\citet{Cuzzi2008} also looked into the effect of internal turbulent pressure on the diffusion of pebble clouds as a limiting factor for gravitational collapse, but they argued that the headwind a collapsing pebble cloud experiences is stronger than the effect of global turbulence with a strength of even $\alpha = 10^{-3}$, thus they neglected the internal diffusion thereafter \citep{Cuzzi2010}.
As they point out in \citet{HartlepCuzzi2020} the untamed headwind in the solar nebula would make the formation of panetesimals as small as 100 km impossible, so they introduced a head wind reduction factor $F_\beta = 1 / 30$ arguing that planetesimal formation may occur in a zonal flow. This falls pretty much along our point of view, yet as they decrease the effect of headwind, the assumption that headwind is stronger than internal diffusion does no longer hold. As we will show later, the inclusion of turbulent diffusion should strongly change the results from \citep{HartlepCuzzi2020}.

As a side remark, our simulations (see Section 4) contain the full un-reduced headwind, which is an essential part of the streaming instability we study. So we see no severe impact of this headwind on the collapse, which is entirely diffusion controlled. 

\begin{figure*}
	\begin{center}
		\includegraphics[width=1.0\textwidth]{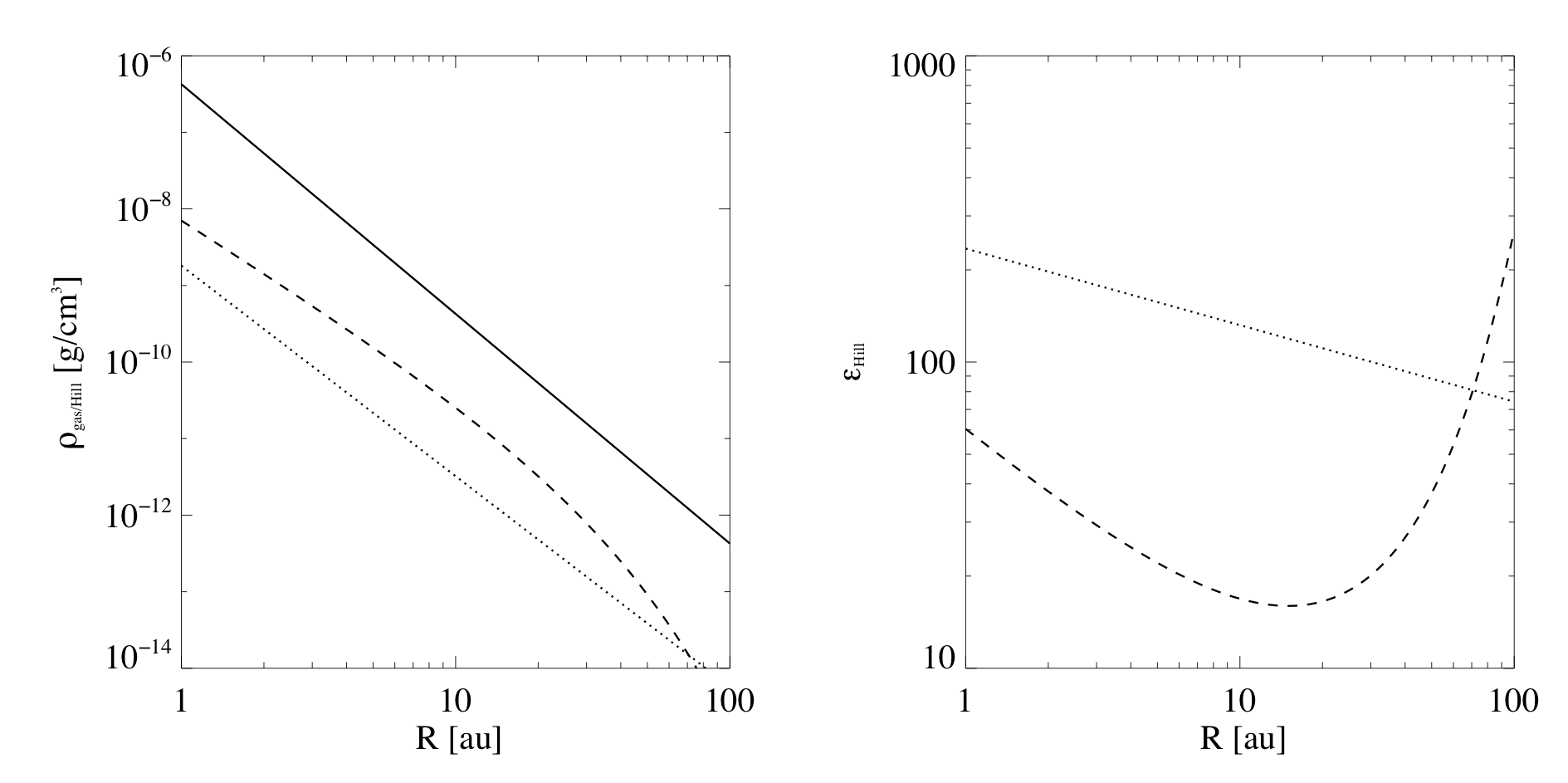}
	\caption{A comparison of gas density and Hill density in models of the solar nebula. Left side: Gas midplane density as in the Minimum Mass Solar Nebula of \citet{Hayashi1981} (dotted line) and of the Most Appealing Solar Nebula in \citet{Lenz2020} (dashed line) in comparison to the Hill density for a solar mass type star (solid line). Right side: The resulting dust to \corrected{gas} ratios $\varepsilon_\textrm{Hill}$ at reaching Hill density in the both cases. (dotted = Hayashi; dashed = Lenz)}
	\label{f:densities}
	\end{center}
\end{figure*}

\subsection{A time scale argument}

The particle cloud collapse criteria that we want to derive resembles the Jeans mass \citep{Jeans1902} criteria, which is the lowest mass threshold for a molecular cloud core to form stars. It is defined by questioning if internal gas pressure can be overcome by self-gravity, resulting in the cloud collapsing under its own weight. Thus, the Jeans mass is a function of cloud density and temperature. The situation for planetesimal formation, starting from a self-gravitating pebble cloud, is similar, but not the same. The self-gravity of the particle cloud has to overcome two opposing effects: On one hand, the cloud has to withstand the tidal shear forces exerted by the central star. This is a force strong enough to disrupt comets, like Shoemaker-Levy \citep{Asphaug1996} in 1993 during its encounter with Jupiter. On the other hand, gravity has to overcome diffusion from inherent turbulent motions of the gas-dust-mixture \citep{Shariff2015}. Hence, objects smaller than the derived critical size can not collapse, but are diffused by gas turbulence, an effect neglected in the original works of planetesimal formation \citep{Safronov1969,GoldreichWard1973}, which only considered gas free r.m.s.\ velocities among the particles.

We state two criteria for a gravitational collapse to occur: 

\subsubsection{First Criterion} The density of a particle cloud has to be larger than the critical density $\rho_{\rm Hill}$ (\citet{Johansen2014} and Appendix B) that allows the cloud to withstand tidal shear while in orbit around a star of mass $M$ at a distance of $R$. 
\begin{equation}
\rho_{\rm Hill} = \frac{9}{4 \pi}\frac{M}{R^3}.
\end{equation}%
In Fig.\ \ref{f:densities} we plot the Hill density as function of distance around a solar mass star. We compare the value to the Minimum Mass Solar Nebula (MMSN) profile \citep{Hayashi1981} density profile of the gas (1700 $g/cm^3$ at 1 au and a radial slope of $-1.5$) and find that the Hill density exceeds the gas density by a factor of about $200-80$ depending on the location in the disk. 

We also compare the Hill density to a more modern approach of a nebula that would be able to form the solar system at least in terms of planetesimal distribution \citep{Lenz2020}. 
In that paper the authors apply the aforementioned pebble flux regulated planetesimal formation rate and constrain initial disk parameters, like disk mass, radial slope, plus the best suited $\alpha$ viscosity,
to produce an initial planetesimal distribution for the solar nebula, that fits all known constraints to form planets in the current mixed pebble and planetesimal accretion scenario to form terrestrial planets \citep{Walsh2011} and cores \citep{raymond2017} and for the dynamic evolution as in the Nice model \citep{Morbidelli2007,Levison2011} among others.

In the Most Appealing Solar Nebula (MASN) \citep{Lenz2020} the disk is more massive $0.1 M_\odot$ and the local gas density is larger than in the MMSN case. The MASN is also subject to viscous evolution, thus the gas surface density is shallower than in the MMSN case. The MMSN was always to steep to be explained by viscous evolution. Also disks in star forming regions observed in the sub-millimeter \citep{Andrews2010} show radial density slopes as shallow as predicted by viscous modelling. 

The MASN has an exponential cut-off radius at $20$ au, which was result of fitting the constrain of a low mass planetesimal disk for the Nice Model \citep{Morbidelli2007}. As also shown in Fig.\ \ref{f:densities}, the dust to gas ratio in the MASN at reaching Hill density $\varepsilon_\textrm{Hill}$ is initially probably between $10$ and $100$, which is according to \citet{Lenz2020} when most planetesimals in terms of bulk mass will form. 

Streaming instability is still driving turbulence diffusion even at dust to gas ratios up to $1000$ as shown in \citet{Schreiber2018} and we will come back to this issue when discussing planetesimal sizes as a function of local dust to gas ratios.
\subsubsection{Second Criterion:} If the critical density $\rhoHill$ is reached, the gravity at the cloud surface still has to overcome the turbulent diffusion, characterised by the diffusion coefficient $D$. 
Now, in order to form a planetesimal, a particle cloud has to contract faster than the diffusion can disperse it.
If the particles were large enough to decouple completely from the gas, then the collapse of a cloud at density $\rho_{\rm Hill}$ would occur on the free-fall time $\tau_{\rm ff}$. 
\begin{equation}
\tau_{\rm ff} = \sqrt{\frac{3 \pi}{32 G \rho_{\rm Hill}}} = \sqrt{\frac{\pi^2}{24 \Omega^2}} =  0.64 \Omega^{-1}.
\label{eq:collapse}
\end{equation}
In the classical work of gravitational formation of planetesimals \citep{Safronov1969, GoldreichWard1973} gas drag is neglected and the collapse is controlled by the r.m.s.\ velocity of the colliding particles and the relevant coefficient of restitution.
While it is certainly the case that during the collapse of the pebble cloud collisions among the pebbles will eventually dominate \citep{Nesvorny2010} this is not the case for the onset of collapse \citep{Wahlberg2017b} and certainly not before the collapse, i.e.\ as long as diffusion can prevent the collapse. As derived in the appendix \ref{sec:collisions}, we can safely ignore pebble collisions in our study as the mean free path for pebbles is larger than the scales of the considered pebble clouds, which is independent from the actual r.m.s.\ velocity. We also derive a critical pebble cloud mass, expressed as a equivalent diameter following the estimates in \citep{Nesvorny2010}\footnote{Note that \citet{Nesvorny2010} derive a radius, whereas we derive a diameter. For details see appendix.}, which we plot in Fig. \ref{Fig:REQ}. We find that only pebble clouds of significant larger mass (respectively equivalent diameter) than the sizes derived in our paper are subject to collisions during the onset of collapse. 

\begin{figure}
\begin{center}
\includegraphics[width=1.0\linewidth]{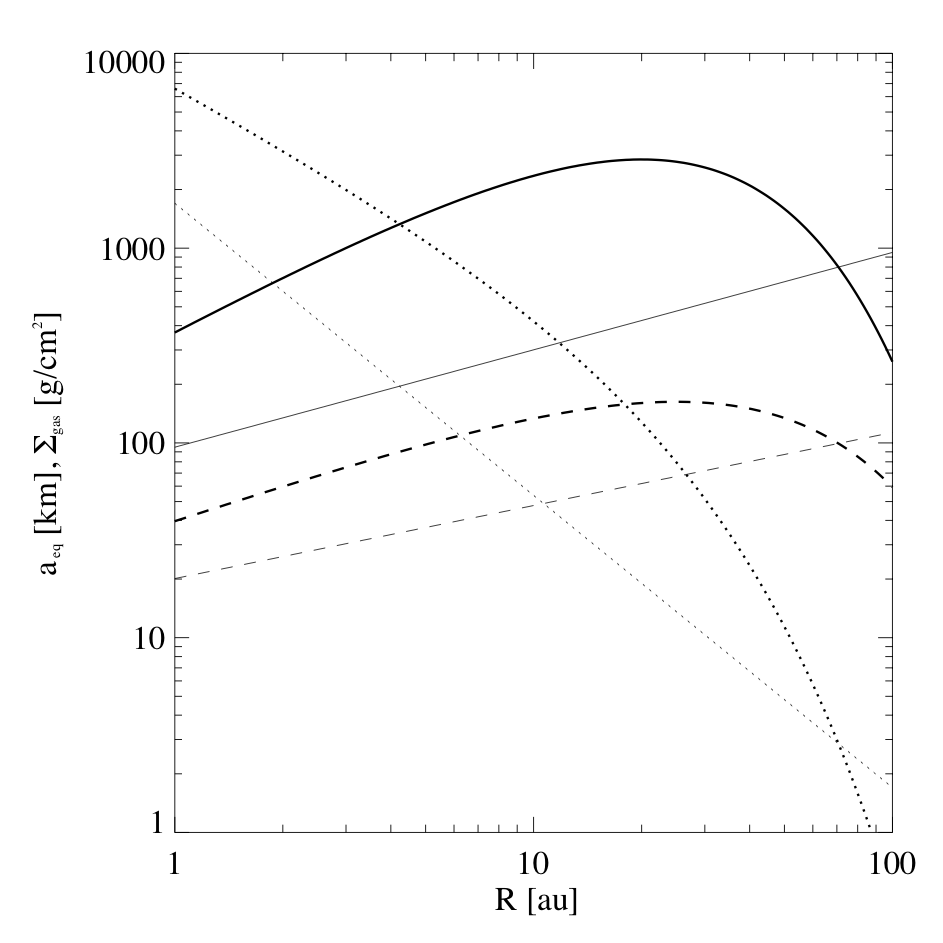}
\caption{Critical equivalent diameter for a pebble cloud (solid line) to find collision time $\tau_\textrm{coll}$ to equal gas friction time $\tauS$ (thick = MASN \citep{Lenz2020}; thin = MMSN \citep{Hayashi1981}). Lower mass pebble clouds at Hill density are dominated by gas friction and larger ones by collisions. See derivation in the appendix. The dotted lines indicate the individual surface density profiles for both disk models. We also add the equivalent diameter of gravitational unstable pebble clouds as dashed lines, that we will derive in Section \ref{sec:AEQ} see Fig. \ref{fig:aeq2}}.%
\label{Fig:REQ} %
\end{center}
\end{figure} %
   
\citet{Nesvorny2010} compare the friction time to the collision time, for the onset of collapse of a pebble cloud forming Kuiper Belt binaries of equivalent size of 500 km diameter and find the collision time to be an order of magnitude shorter than the friction time and therefore neglect gas-drag.
They assume the pebble cloud to be already virialized and use the virial velocity as r.m.s.\ velocity, 
yet \citet{Wahlberg2017b} argue that the contraction speed of the pebbled cloud in the presence of gas drag is initially larger than the r.m.s.\ speed, thus gas friction again wins over collisions for the onset collapse.

Nevertheless, we also added an estimate for the ratio between collision and friction time for our simulations to the appendix \ref{sec:collisions} and determine critical pebble cloud masses (respectively equivalent diameter $a_\mathrm{eq}$) for the MMSN and MASN (see Fig.\ \ref{Fig:REQ}),
and find agreement with \citet{Nesvorny2010} about 500 km that our derived pebble cloud masses are always too small to be dominated by collisions for the onset of collapse.

\correctedold{Translating a given Stokes number into physical grain or pebble diameter results in a range of sizes from sub-\mm~to several \cm, depending on their porosity and external conditions like the gas surface density of the disk}. In other words a massive cob web, a large snow flake and a small marble can have very different masses and sizes, yet can still share a common friction time, which is all what matters for both pebble definition and planetesimal formation.
\begin{figure}
    \centering
    \includegraphics[width=1.0\linewidth]{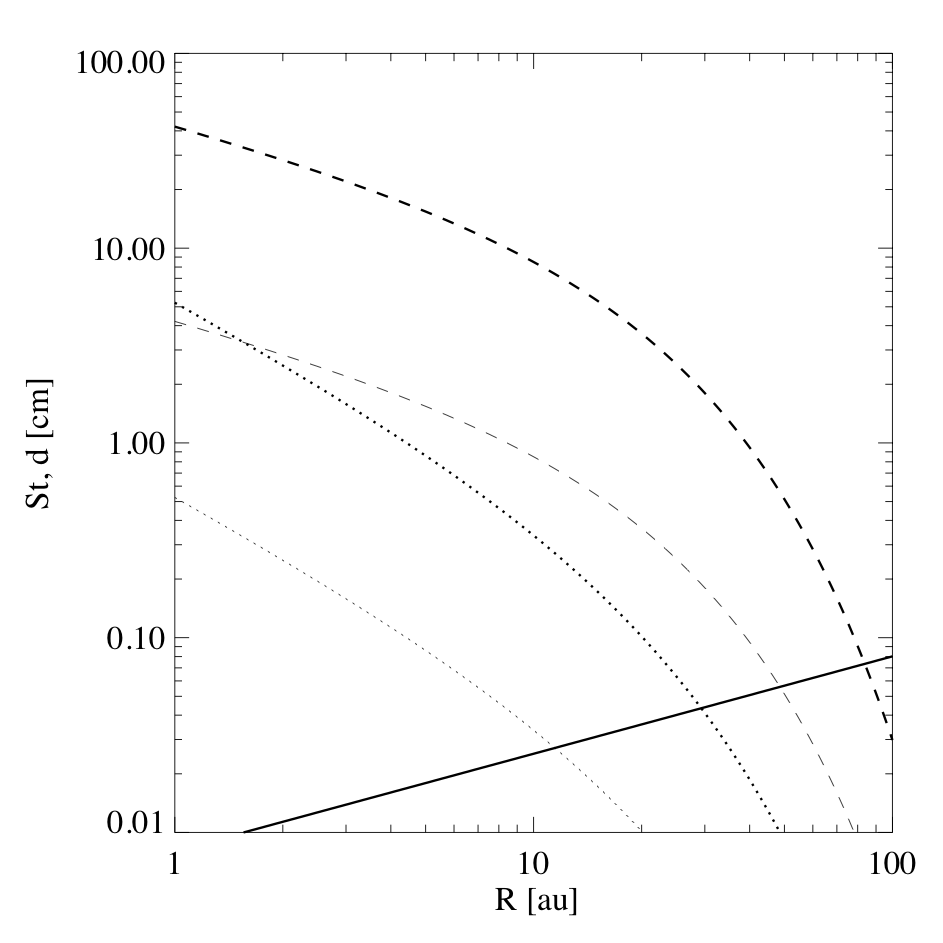}
    \caption{Stokes Number as function of radius for the MASN model \citep{Lenz2020} (solid line). Dashed lines are the diameters $d$ of the pebbles for the initial gas mass (thick dashed line) and for a reduced mass to $10\%$ of the initial value (dashed thin line). The dotted lines are pebble diameter for pebbles too small to be forming planetesimals $\St= \num{1e-3}$, i.e.\ material that will slowly be accreted to the surfaces of already formed planetesimals. For the initial gas mass (thick dotted line) this is cm material at 2-3 au and after about a Million years (lower gas mass), mm sized material.}
    \label{fig:dust}
\end{figure}
Interestingly \correctedold{the Stokes-numbers of pebbles around} 2 au in our models of planetesimal formation in the solar nebula \citep{Lenz_2019, Lenz2020} (see Fig.\ \ref{fig:dust}) are \correctedold{corresponding to solid marbles of a several cm. Thus if we assume those big pebbles to be incorporated into planetesimals and material smaller than $\St = 0.01$ to be largely excluded from this process \citep{Lenz_2019}, then we would expect $\St \approx 0.001$ material to be left behind. In the early more gas rich stages of the solar nebula this Stokes-Number would correspond to roughly 1 cm sized material like calcium-aluminum-rich inclusions (CAIs) \citep{2012Sci...338..651C} and at later stages to mm sized pebbles, like the typical chondrules either before or after the flash heating \cite{2019LPI....50.2919D}.
In that paradigm CAIs and chondrules would be the pebbles that were just a bit too small to be effectively involved in planetesimal formation and the last ones to be subject to pebble accretion.}
\correctedold{Following \citet{Wahlberg2017} pebbles of $\St < 1$ feel the friction with the gas as the dominant effect controlling the onset of collapse and the actual contraction time $\tau_{\rm c}$
will become longer than the free fall time} \citep{Shariff2015}. In the case of the friction time being shorter than the collapse time, the contraction time is inversely proportional to the friction time $\tau_{\rm f}$, as derived in the appendix C.
For $\St < 0.1$ the contraction time is: 
\begin{equation}
\label{eq:contractionTime}
\tau_{\rm c} =  \frac{8}{3 \pi^2} \frac{\tau_{\rm ff}^2 }{\tau_{\rm f} }
\end{equation}
For $\St > 0.1$ the contraction time approaches the free-fall time. 
As the free-fall time enters this 
expression as squared the contraction time is
inversely proportional to the actual particle density of the shrinking cloud 
\begin{equation}
\tau_{\rm c}= \frac{1}{9 \St} \frac{\rho_{\rm Hill}}{\rho}\, \Omega^{-1} = \frac{1}{9 \St} \frac{r(t)^3}{r_0^3}\, \Omega^{-1},
\end{equation} 
which means that once a cloud of radius $r_0$ is able to contract, the process accelerates with the shrinking cloud radius $\propto r(t)^{3}$.
If the contracting particle cloud of radius $r$ is subject to turbulent motion then the cloud is diffused on the typical timescale of 
\begin{equation}
\label{eq:diffusionTime}
\tauDiff = \frac{r^2}{D}=   \frac{1}{\delta} \left(\frac{r} {H}\right)^2\Omega^{-1},
\end{equation}
where we use the dimensionless diffusivity $\delta$ by scaling $D = \delta H c_s$ with the vertical disk extent $H$ (aka pressure scale height) and the speed of sound $c_s = H \Omega$.
\begin{figure*}
\begin{center}
\includegraphics[width = \textwidth, angle=0]{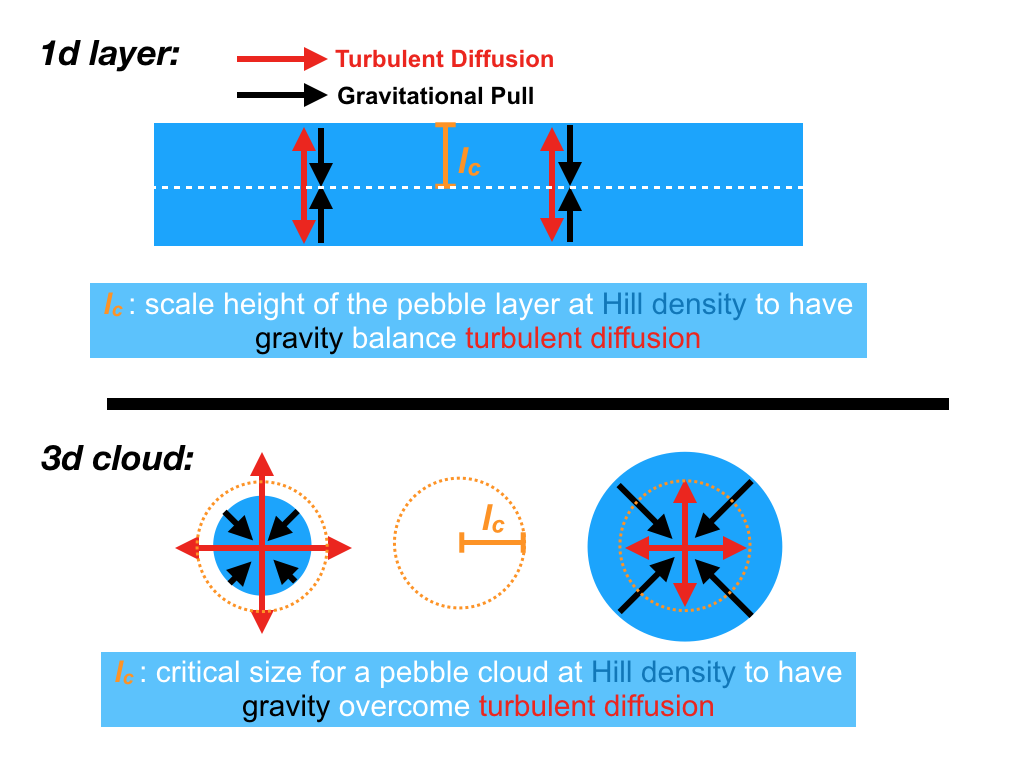}
\caption{Critical Length scale scheme: The competition between the Self Gravity and turbulent diffusion defines a critical length scale $l_{\rm c}$. Upper panel: vertical sedimentation up to the Hill density in the midplane leads to a layer that cannot contract anymore, because in this 1D - plan parallel case eventually diffusion is stronger than self-gravity (see appendix). The scale height of such a layer with Hill density in the center is given by $l_{\rm c}$. Lower panel: The same $l_{\rm c}$ defines the critical length in the 3D case, when a pebble cloud at Hill density cannot be stabilized by internal diffusion anymore (see text). Smaller clouds get dispersed (left), larger ones collapse (right). \label{fig:1}}
\end{center}
\end{figure*}
Regardless of whether the diffusivity measured in $\delta$ is due to large scale gas turbulence in the solar nebula or solely generated by the streaming instability, or some combination of both, this expression does hold.
We will discuss this issue below when we use in real numbers to estimate the resulting mass for planetesimals.

The diffusion time scales with $r^2$ but the contraction time with $r^3$, thus the contraction will always win, once started. Once a pebble cloud goes beyond a critical mass, the contraction can not be halted by diffusion anymore. This is the same effect as the collapse of an isothermal sphere of gas, which cannot be stabilized by the increase of its internal pressure.

Comparing the two timescales of diffusion (Eq.~\ref{eq:diffusionTime}) and contraction (Eq.~\ref{eq:contractionTime}) by setting $\tau_{\rm c} = \tauDiff$, we can derive a critical \correctedold{(minimal)} cloud radius $r = \lcrit$ for a 
pebble cloud at Hill density to withstand internal diffusion and allow for contraction at 
\begin{equation}
\label{eq:collapseCrit}
\lcrit = \frac{1}{3} \sqrt{\frac{\delta}{\St}} H\, .
\end{equation}

This expression can be understood as a critical length for planetesimal formation, similar to the Jeans length in star formation. Clumps of less than the critical density or of radius less than $\lcrit$ cannot collapse, but larger or more massive clumps will do  (See Fig. \ref{fig:1}).
Moreover, cloud collapse will always set in, once this border of stability is reached, thus larger or more massive pebble clouds are less likely to form, if accumulation takes longer than the gravitational collapse, thus limiting planetesimal sizes.

The classical Jeans length is derived for a homogeneous density distribution, asking when an infinitesimal density perturbation will start to grow and collapse. A derivation of this smallest linear unstable wavelength for pebbles in a Toomre instability fashion gives $\lambda = 2 \pi \lcrit$ (see appendix).
This means that in the following simulations, in which our simulation size $L$ never covers $L = 2 \pi \lcrit $, the streaming instability triggered the formation of the clumps and not linear growing self gravity modes, because they not fit in our simulation domain. 
The $\lcrit$ criterion is \correctedold{therefore} describing the stability of non-linear density perturbations as created by the streaming instability or any other concentrating effect, when local gas turbulence has to be considered.

\subsection{Particle layer scale height}

The length scale $\lcrit$ is a more general quantity than we mentioned so far. If we ask for the vertical scale height of the dust sub disk in the solar nebula at reaching the Hill density in the midplane, i.e.\ in equilibrium between vertical diffusion and sedimentation dominated by self gravity (See Fig.\ref{fig:1}), we find the functional dependency 
\begin{equation}
    \rho_d(z) = \rho_H \cosh^{-2} \left(- \frac{z}{\sqrt{2} \lcrit}\right) \approx \rho_H e^{- \frac{z^2}{2 \lcrit^2}},
\end{equation}
which for values in $z$ up to one pressure scale height can be approximated with the usual Gaussian distribution of density around the mid-plan with an error of less than $4\%$ (see Fig.\ \ref{f:layer}). This means that our critical length-scale is simultaneously the "pressure" scale height of the self gravitating pebble accumulations. Without self gravity the scale height of particles $h_p$ \citep{Dubrulle1995} would be three times larger,
\begin{equation}
\label{eq:collapseCrit2}
h_p = \sqrt{\frac{\delta}{\St}} H = 3 \lcrit.
\end{equation}
\correctedold{With increasing mass of the pebble layer it will also become thinner and thus over-proportionally denser.}

\section{Bringing our prediction to a numerical test}
\begin{figure*}
\gridline{
          \fig{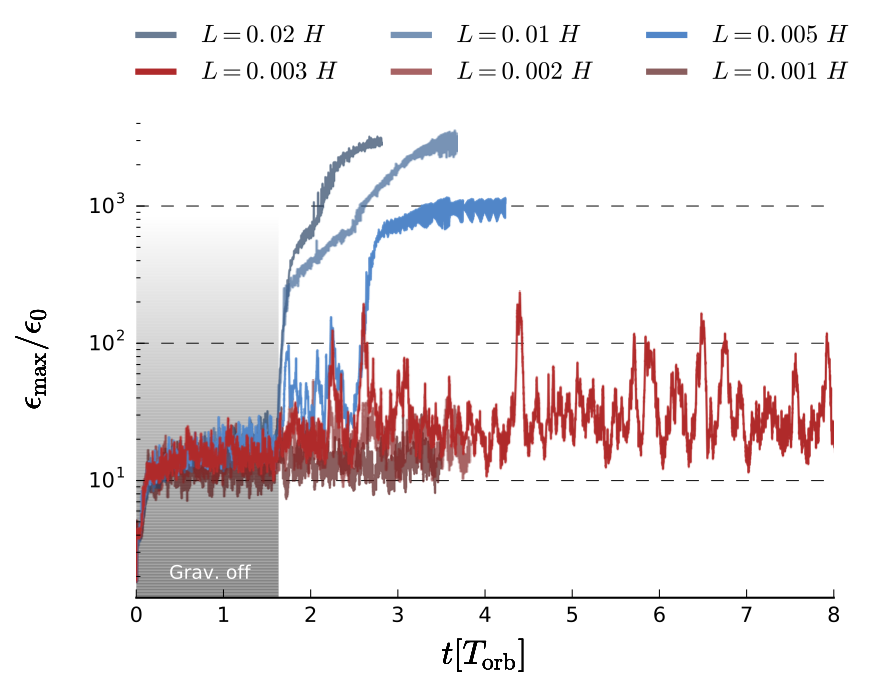}{0.5\textwidth}{(a): \textbf{A} runs with \textbf{$\St=0.1$}}}
\gridline{
          \fig{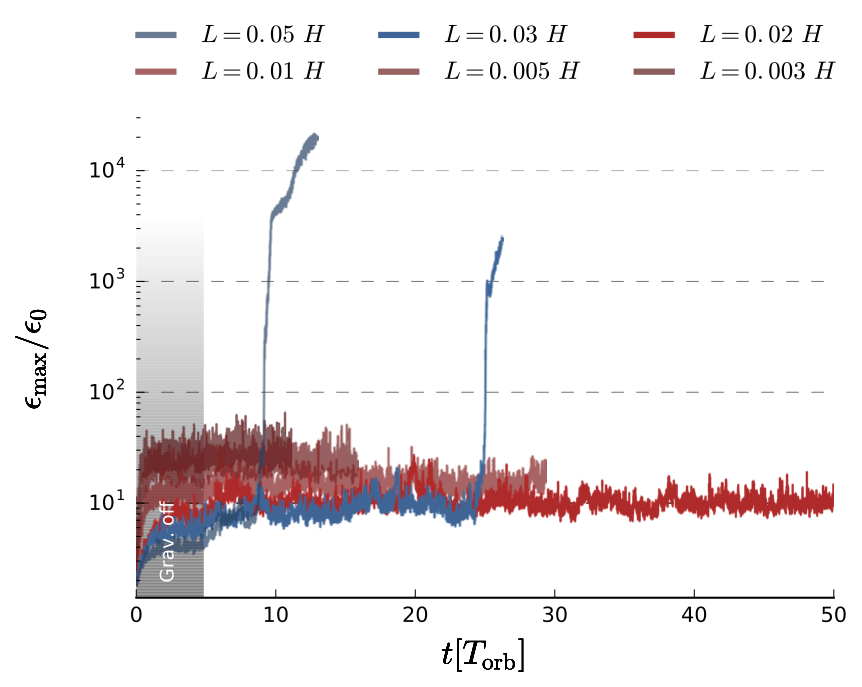}{0.5\textwidth}{(b): \textbf{B} runs with \textbf{$\St=0.01$}}
          }
\caption{Time evolution of maximum dust-to-gas ratio. 
Simulations with $L> 2 \lcrit$ are colored in blue, smaller simulations with $L< 2 \lcrit$ in red. Particles self-gravity is turned on at $t=1.59 T_{\rm orb}$ (\texttt{A} runs) and $t=4.8 T_{\rm orb}$ (\texttt{B} runs). Since a smaller Stokes number means longer collapse time (see App.~\ref{sec:collapseTime}) the simulations with $\St=0.01$ takes longer to collapse. The border cases just below the instability criterion (bright red) have been running the longest to show the validity of this criterion. Additionally to the collapse, this is when the $\varepsilon_\mathrm{max}$ increases by orders of magnitude within a short time, one sees post-formation growth and merging events of this fragmented objects. The \texttt{B} runs show an additional increase in maximum solid concentration due to the streaming instability but still our criterion holds.}
\label{fig:timeseries}%
 %
\end{figure*}

To test the derived $\lcrit$-criterion we perform shearing box simulations of self-gravity induced collapse in a gas and dust mixture starting from fully developed streaming instability turbulence.
\correctedold{The idea is that only in a simulation with box size $L$ in which a cloud of diameter $2 l_c$ would fit can lead to a collapse.
Thus for the numerical simulations the collapse criterion is:
\begin{equation}
    L > 2 l_c
\end{equation}
We perform simulations for two particle sizes that is for}
for $\St=0.1$ and $\St=0.01$ particles (Fig.\ \ref{fig:2}) and choosing an average dust to gas ratio of $\varepsilon_0 = 3$, suggesting that trapping and sedimentation have already achieved this level of local dust concentration. We used the Pencil Code in setups based on \citet{Schreiber2018} in a radial-azimuthal setup to \correctedold{save} computation time. See table \ref{tab:Ste-1Runs} for the simulation parameters. All simulations have the same numerical resolution of $256^2$ grid cells and same initial dust-to-gas ratio $\epsInit=3$. The physical domain size $L$ is altered around the predicted critical length scale $2 \lcrit$. Two simulations around $2 \lcrit$ with $\St=0.1$ are additionally altered in gas pressure gradient $\eta$. Maximum dust density timeseries can be found in Fig.~\ref{fig:timeseries}. The z-dimension has only one grid cell. Table \ref{tab:simResults} gives the measured turbulence and diffusivity including the predicted length-scale $\lcrit$.

Different resolution and different sizes of the box changes the strength of the streaming instability and thus the turbulent diffusion \citep{Schreiber2018}. Therefore each setup has a different critical $l_c$, even for the same pressure gradient, dust to gas ratio and Stokes number. \correctedold{As a consequence we} do not determine here \correctedold{the ultimate} value for $\delta$ for streaming instability in general, for which one would need global 3D high resolution studies with a wide range of Stokes numbers, but we focus on testing the validity of the $2 l_c < L$ \correctedold{criterion by varying the box size $L$. An alternative method is to keep $L$ fixed, but to alter the pressure gradient,} which we have shown by means of additional simulations (\texttt{Ae3L0005lp, Ae3L0005hp, Ae3L0003lp}).

\begin{deluxetable}{llllr}
	\tablecaption{Overview over all simulations:
	The name indicates the Stokes Number: \texttt{A}: St = 0.1 and \texttt{B}: St = 0.01. The rest of the name refers to the initial dust to gas ratio (always the same) and to Domain size $L$ in the following column, grid spacing $d_\mathrm{x,y}$, gas sub-Keplerianicity $\eta$ and maximum simulation run time in orbits. Self-gravity is turned on at $T=1.59T_{\rm orb}$ in the \texttt{A} runs and at $T=4.77\torb$ for the \texttt{B} runs. Additional simulations where performed with variation in the pressure gradient $\eta$ by a factor of $2$ (hp = high pressure) or by a factor 
		$\onehalf$ (lp = low pressure). \label{tab:Ste-1Runs}}
\tablehead{Name      & $L_{\rm x}$, $L_{\rm y}$           & $d_{\rm x,y}$ & $\eta$ & $T_{\rm max} \left[T_{\rm orb}\right]$} 
\startdata
	\texttt{Ae3L002}    & $\SI{0.02}{\scaleheight}$  &  \num{7.81E-5}   & 0.05   &                              \num{2.82} \\
	\texttt{Ae3L001}    & $\SI{0.01}{\scaleheight}$  &  \num{3.91E-5}   & 0.05   &                              \num{3.67} \\
	\texttt{Ae3L0005}   & $\SI{0.005}{\scaleheight}$ &  \num{1.95E-5}   & 0.05   &                              \num{4.24} \\
	\texttt{Ae3L0005lp} & $\SI{0.005}{\scaleheight}$ &  \num{1.95E-5}   & 0.025  &                             \num{13.06} \\
	\texttt{Ae3L0005hp} & $\SI{0.005}{\scaleheight}$ &  \num{1.95E-5}   & 0.1    &                             \num{32.78} \\
	\texttt{Ae3L0003}   & $\SI{0.003}{\scaleheight}$ &  \num{1.17E-5}   & 0.05   &                             \num{10.03} \\
	\texttt{Ae3L0003lp} & $\SI{0.003}{\scaleheight}$ &  \num{1.17E-5}   & 0.025  &                             \num{14.47} \\
	\texttt{Ae3L0002}   & $\SI{0.002}{\scaleheight}$ &  \num{7.81E-6}   & 0.05   &                              \num{3.83} \\
	\texttt{Ae3L0001}   & $\SI{0.001}{\scaleheight}$ &  \num{3.91E-6}   & 0.05   &                              \num{3.50} \\ 
	\texttt{Be3L005}    & $\SI{0.05}{\scaleheight}$  &  \num{1.95E-4}   & 0.05   &                              \num{12.57} \\
	\texttt{Be3L003}    & $\SI{0.03}{\scaleheight}$  &  \num{1.17E-4}   & 0.05   &                              \num{26.22} \\
	\texttt{Be3L002}    & $\SI{0.02}{\scaleheight}$  &  \num{7.81E-5}   & 0.05   &                              {\num{50.93}} \\
	\texttt{Be3L001}    & $\SI{0.01}{\scaleheight}$  &  \num{3.91E-5}   & 0.05   &                              {\num{31.83}} \\
	\texttt{Be3L0005}   & $\SI{0.005}{\scaleheight}$ &  \num{1.95E-5}   & 0.05   &                              {\num{16.84}} \\
	\texttt{Be3L0003}   & $\SI{0.003}{\scaleheight}$ &  \num{1.17E-5}   & 0.05   &                              {\num{11.58}} \\ 
	\enddata
\end{deluxetable}

\begin{figure}
\begin{center}
\includegraphics[width=\linewidth]{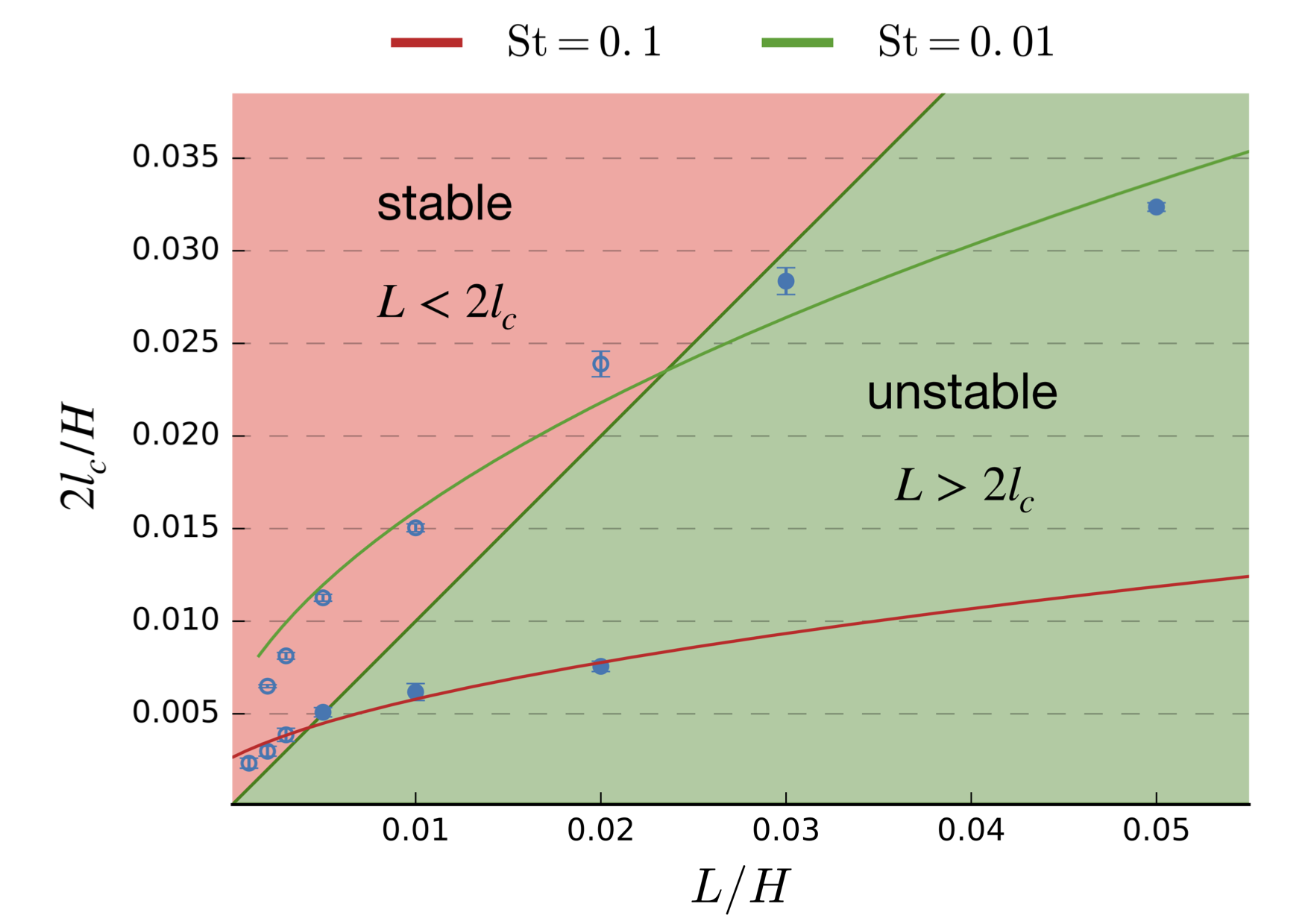}
\caption{Numerical results compared with analytic prediction. With domain size $L$ on the x-axis we plot for two different Stokes numbers, i.e. particle sizes, the correspoding critical length scale $\lcrit$. This scale is determined by measuring the diffusivity of the pure streaming instability before switching on self-gravity. The red region indicates $L < 2 l _{\rm c}$ where no collapse should be possible because the diffusion is too strong whereas in the green region $L > 2 l _{\rm c}$ collapse should occur. We find agreement between our prediction and the simulation results: All simulations with filled symbols did collapse and the ones with open symbols did not. \label{fig:lcrit}}
\end{center}
\end{figure}

\begin{figure*}
\begin{center}
\includegraphics[width = \textwidth, angle=0]{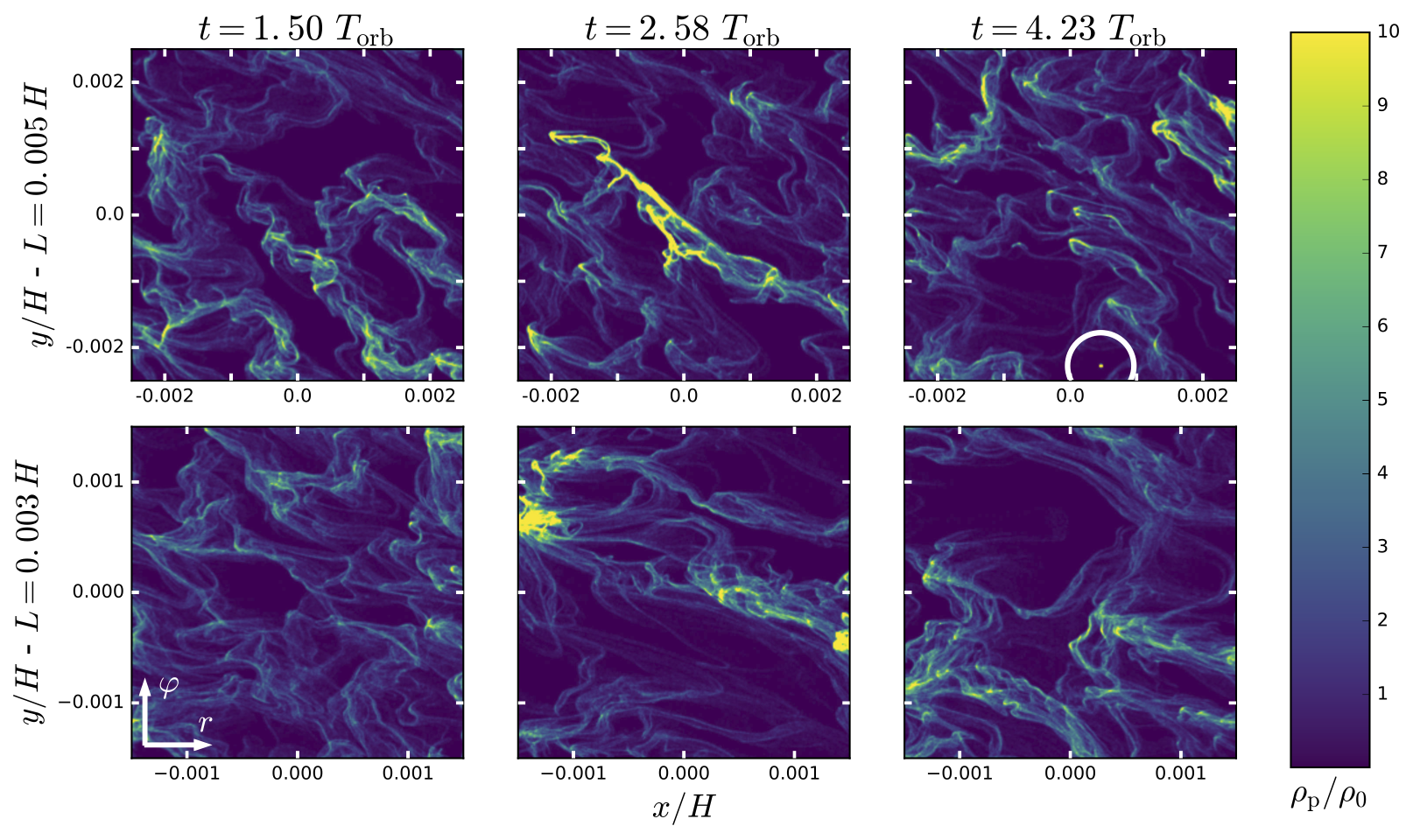}
\caption{Distribution of 0.1 $\St$ particles for the two critical simulations. The two rows compare simulations with only slightly different box-sizes $L$. The upper covers the critical collapse length $L \approx 1.1 \cdot 2 l_{\rm c}$, as given by the underlying turbulent diffusion, but the lower with $L \approx 0.75 \cdot 2 l_{\rm c}$ does not (Fig. 1). 
The left column shows the typical streaming instability pattern without gravity. The central column is taken one orbit after dust self-gravity has been turned on and shows now a gravoturbulent situation where both simulations had a similar large local dust enhancement by a factor of 200. Here, both simulations show an rather elongated filament filling almost the entire domain, but only in the larger simulation this filament can contract against turbulent diffusion and finally collapse. As a result, out of this overdensity only the upper simulation was able to produce a planetesimal, highlighted by a white circle in the right column. All simulations are performed at Hill density. \label{fig:2}}
\end{center}
\end{figure*}

In comparison to other work \citep{Johansen2015,Simon2016}, we were able to ensure to resolve the critical length scale $\lcrit$ by $64$ grid cells or more. We varied the size of the simulation domain, $L$, in a set of models at Hill density and following our prediction, only boxes larger than  $2 \lcrit$ collapsed (Fig.\ \ref{fig:lcrit}), i.e. when a cloud of radius $\lcrit$ would have fit into the box. %
The radial diffusivity $\delta_x$ is measured in the situation of saturated streaming instability, but before gravity is switched on, see Fig.~\ref{fig:diffusion}. In this measurement, the diffusivity increases with simulation domain size $L$, since larger modes of the streaming instability are stronger diffusing particles, which are suppressed in smaller simulation domain sizes. As shown in this paper the simulations \texttt{Ae3L0003}, \texttt{Ae3L0002} and \texttt{Ae3L0001} are the ones not collapsing from our \texttt{A} parameter set. Even \texttt{Ae3L0003} is not collapsing after more than 8 orbits. On this scales, diffusion acts faster than collapse, whereas on scales larger than $L\geq0.005\scaleheight$ planetesimals did form. 

\begin{deluxetable*}{llccccccc}
\tablewidth{0.5\linewidth}
	\tablecaption{Simulation results: \texttt{A} runs with $\St=0.1$ particles and \texttt{B} runs $\St=0.01$. Diffusivities and velocities are measured by tracking the radial position of $10^4$ particles for several orbits and treating it similar to a turbulence driven random walk. Diffusivity and rms-velocities are measured in the non-gravitating fully turbulent situation. The number of planetesimals $N_\mathrm{p}$ is the number of objects we find in our final snapshots. \label{tab:simResults}}
\tablehead{		Name                & $N_\mathrm{p.}$ & $\delta_\mathrm{x}$ & $\Delta\delta_\mathrm{x}$ & $\lcrit$ & $u_\mathrm{rms}$ & $u_\mathrm{rms,x}$ & $v_\mathrm{rms}$ & $v_\mathrm{rms,x}$} 
\startdata
		\texttt{Ae3L002}    & 2               &    \num{1.23E-5}    &       \num{4.80E-8}       & \num{7.40E-3}  &  \num{5.77e-3}   &   \num{6.93e-3}    &  \num{4.02E-3}   &   \num{5.02e-3}    \\
		\texttt{Ae3L001}    & 8               &    \num{8.34E-6}    &       \num{6.94E-8}       & \num{6.09E-3}  &  \num{4.24e-3}   &   \num{6.78e-3}    &  \num{3.05E-3}   &   \num{4.88e-3}    \\
		\texttt{Ae3L0005}   & 1               &    \num{5.86E-6}    &       \num{2.95E-8}       & \num{5.10E-3}  &  \num{3.55e-3}   &   \num{4.39e-3}    &  \num{2.74e-3}   &   \num{3.22e-3}    \\
		\texttt{Ae3L0005lp} & 3               &    \num{2.25e-6}    &       \num{1.19e-6}       & \num{3.16E-3}  &  \num{2.27e-3}   &   \num{2.60e-3}    &  \num{1.79e-3}   &   \num{2.00e-3}    \\
		\texttt{Ae3L0005hp} & 0               &    \num{1.48e-5}    &       \num{9.09e-6}       & \num{8.12E-3}  &  \num{6.20e-3}   &   \num{8.08e-3}    &  \num{4.76e-3}   &   \num{6.44e-3}    \\
		\texttt{Ae3L0003}   & 0               &    \num{2.26E-6}    &       \num{1.36E-8}       & \num{3.17E-3}  &  \num{2.55E-3}   &   \num{4.20e-3}    &  \num{1.71e-3}   &   \num{2.96e-3}    \\
		\texttt{Ae3L0003lp} & 1               &    \num{1.04e-6}    &       \num{5.61e-7}       & \num{2.15e-3}  &  \num{1.59e-3}   &   \num{2.52e-3}    &  \num{1.23e-3}   &   \num{2.07e-3}    \\
		\texttt{Ae3L0002}   & 0               &    \num{2.00E-6}    &       \num{1.58E-8}       & \num{2.98E-3}  &  \num{1.62E-3}   &   \num{2.61e-3}    &  \num{1.34e-3}   &   \num{1.82e-3}    \\
		\texttt{Ae3L0001}   & 0               &    \num{1.31E-6}    &       \num{8.04E-9}       & \num{2.41E-3}  &  \num{1.79E-3}   &   \num{4.56e-3}    &  \num{0.88e-3}   &   \num{2.84e-3}    \\ 
		\texttt{Be3L005}    & 2               &   \num{2.36e-05}    &      \num{8.25e-06}       & \num{3.24E-2} &  \num{5.68E-3}   &   \num{6.81E-3}    &  \num{5.53E-3}   &   \num{6.61E-3}    \\
		\texttt{Be3L003}    & 1               &   \num{1.81e-05}    &      \num{9.23e-06}       & \num{2.84E-2} &  \num{4.27E-3}   &   \num{5.05E-3}    &  \num{4.07E-3}   &   \num{4.83E-3}    \\
		\texttt{Be3L002}    & 0               &   \num{1.28e-05}    &      \num{7.40e-06}       & \num{2.39E-2} &  \num{3.97E-3}   &   \num{4.91E-3}    &  \num{3.75E-3}   &   \num{4.67E-3}    \\
		\texttt{Be3L001}    & 0               &   \num{5.09e-06}    &      \num{1.44e-06}       & \num{1.50E-2} &  \num{3.44E-3}   &   \num{3.54E-3}    &  \num{3.22E-3}   &   \num{3.32E-3}    \\
		\texttt{Be3L0005}   & 0               &   \num{2.85e-06}    &      \num{9.35e-07}       & \num{1.13E-2} &  \num{3.54E-3}   &   \num{3.09E-3}    &  \num{2.40E-3}   &   \num{2.88E-3}    \\
		\texttt{Be3L0003}   & 0               &   \num{1.49e-06}    &      \num{6.58e-07}       & \num{8.13E-3}  &  \num{3.36E-3}   &   \num{2.77E-3}    &  \num{2.56E-3}   &   \num{2.46E-3}
\enddata
\end{deluxetable*}%

As predicted, the run \texttt{Ae3L0005} is the smallest simulation still being capable to produce a planetesimal and in fact there is only one forming. For the next largest simulation \texttt{Ae3L001} we could count 8 bound objects of different appearing size, varying by a only a factor of 2 in size. Two of them are in a bound binary system, see video 2 in our online material. The \texttt{Ae3L002} simulation produces also the formation of several planetesimals, which start colliding and merging, thus only two planetesimals survive. \correctedold{This collision and merging has to be taken with a grain of salt, as we do not allow our pebble clouds to contract to solid density. We refer to dedicated simulations of cloud collapse as performed by \citet{Nesvorny2010}.}

All rms-velocities are measured in a non-gravitating fully SI turbulent snapshot. Gas rms-velocities are measured by using the grid data, particle rms-velocities by using the complete particle data set. Diffusivities are calculated by tracking a set of $10^4$ particles (see appendix \ref{sec:determinDiff}). The measured values are summarized in \Tab{tab:simResults}.

\begin{figure*}
\centering
\includegraphics[width=\linewidth]{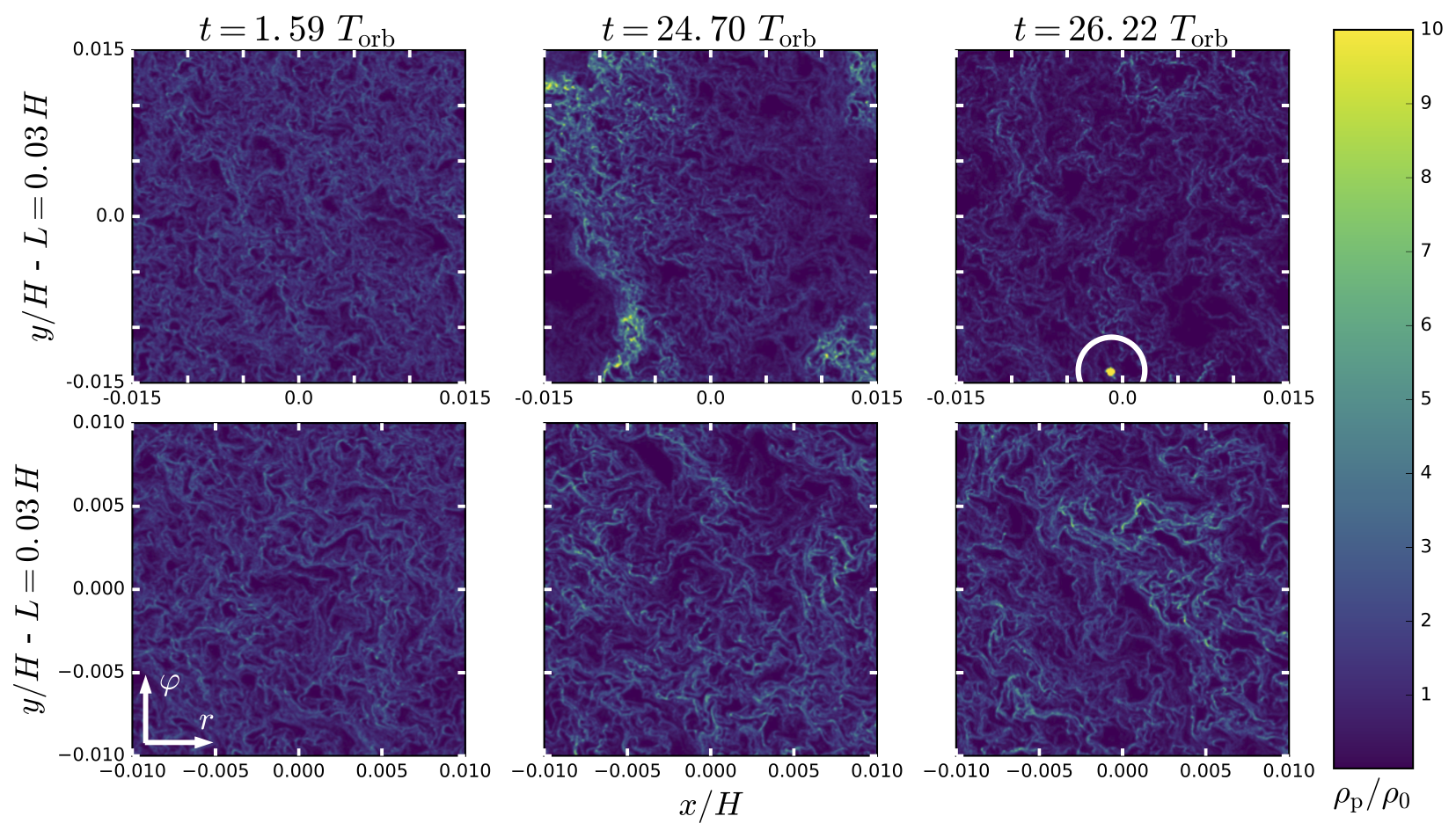}
\caption{\textbf{Distribution of \SI{0.01}{\St} particles for \correctedold{two simulations} around the critical value of $L = 2 l_c$. One is slightly larger and the other one slightly smaller than the critical length. Thus} the two rows compare simulations with only slightly different box-sizes $L$. The upper covers the critical collapse length $L \approx 1.06 \cdot 2 l_\mathrm{c}$ (see Figure \ref{fig:lcrit}), as given by the underlying turbulent diffusion, but the lower with $L \approx 0.84 \cdot 2 l_\mathrm{c}$ does not. The left column shows the typical streaming instability pattern without gravity. The central column is taken one orbit after dust self-gravity has been turned on and shows now a gravoturbulent situation where both simulations had a similar large local dust enhancement by a factor of 200. Here, only the larger simulations shows an elongated filament filling almost the entire domain, because the smaller simulation had produced its filament already at $T=20~T_{\rm orb}$, that got diffused away (see Movie 3 and Figure \ref{fig:timeseries}). As a result, out of this overdensity only the upper simulation was able to produce a planetesimal, highlighted by a white circle in the right column. All simulations are performed at Hill density, i.e. $f=1$.
\label{fig:n_plot_slices_all}}%
\end{figure*}

From the maximum dust density time series for the models using $\St = 0.1$ (Fig.\ \ref{fig:timeseries}) one finds that with decreasing box size it takes longer to form a planetesimal (blue lines), even so diffusivity was getting weaker in those runs. When crossing the border of stability $2\lcrit$ planetesimal formation stalls and the particle cloud remains in a turbulent state. Same effect happened when we altered the pressure gradient (see Fig. 12) and also for the $\St= 0.01$ particles (see Table 3 and Fig. \ref{fig:n_plot_slices_all}). 
Thus, we  find that our criterion reliably predicts the outcome of our numerical experiments for different Stokes numbers and pressure gradients. 
Please see also our online content: Movie 1: \url{https://youtu.be/gkHiluqH8HY} compares simulations \texttt{Ae3L0005} and \texttt{Ae3L0005}. Both use $\St= 0.1$ particles, but only the larger box shows collapse and planetesimal formation.
In Movie 2: \url{https://youtu.be/nA87-9_trUc} we show the evolution of $\St= 0.1$ pebbles for all 6 different box sizes in Table 1, and in Movie 3 we shows the same for the 10 times smaller particles with $\St= 0.01$ \url{https://youtu.be/CCywDPKVU8w}. One clearly sees that for the smaller pebbles the contraction into planetesimals takes longer, as expected. But most importantly, as can be read from Table 3, our collapse criterion always gave the right prediction on whether a simulation would lead to collapse or not.

As also can be seen from Table 3, each simulation found a different critical length scale $l_c$ ranging for the cases that lead to collapse. For $\St= 0.1$ particles that was $l_c = 7.4 - 5.1 \times 10^{-3}$ where as smaller Stokes numbers $\St= 0.01$ had $l_c = 3.2 - 2.8 \times 10^{-2}$. The first trend is that smaller Stokes numbers lead to larger planetesimals, as expected, yet also the diffusivity apparently changed, thus the $\St= 0.01$ unstable cloud was not 10 but only about 5 times larger that the average $\St= 0.1$ pebble cloud. This has to be taken with a grain of salt, note that we only did 2D simulations of a pretty restricted local size to test the $l_c$ criterion. In more global simulations one finds stronger diffusion and interestingly in some cases even a scaling of $\delta \sim \St$, which leads to an $l_c$ independent of an explicit $\delta $ and $ \St$ \citep{Schreiber2018}. More investigations on $\delta $ as a function of $ \St$ in more global setups are desperately needed to constrain $l_c$ directly on (a range of) particle sizes, radial pressure gradient and local dust load in the disk. 
Still our result holds: If you run a turbulent particle and gas simulation for a certain Stokes number and determine the diffusivity before turning on self gravity, then our $L > 2 l_c$ criterion can tell you whether collapse and planetesimal formation will occur.
\subsection{Varying the pressure gradient} %
We added three additional runs around the transition zone from collapse to stability \texttt{Ae3L0005lp}, \texttt{Ae3L0005hp}, \texttt{Ae3L0003lp} by keeping the box-size and resolution as in the \texttt{Ae3L0005} and \texttt{Ae3L0003} runs, but changing the radial pressure gradient. See Figure \ref{fig:lcritA}. The triangle pointing upward indicates a model that was collapsing beforehand, but did not do so if the pressure gradient is doubled, because of stronger turbulence. Reversely, the downward pointing triangles indicate models with a reduced pressure gradient by a factor of two which both collapsed. %
\subsection{Varying the initial dust to gas ratio} %
We also added one additional run by keeping the box-size and resolution of \texttt{Be3L003} but changing the dust-to-gas ratio to $\varepsilon=10$ \texttt{Be3L003e10} (see Figure \ref{fig:lcritA}), indicated with the downward pointing triangle. As expected this run did collapse and demonstrates that increasing the dust load locally will decrease the diffusivity and hence decrease $\lcrit$. %

So different global pressure gradients, different Stokes numbers, and different dust to gas ratios at reaching the Hill density will result in different critical masses for the pebble cloud to undergo collapse. Our $l_c$ and $m_c$ criterion was able to predict all simulation outcomes.

\section{Characteristic planetesimal masses and sizes in the solar nebula.\label{sec:AEQ}}
It makes sense to ask how much mass is in the gravitational pebble cloud of radius $l_c$.
This characteristic mass is the initial condition for the further collapse into one or multiple planetesimals. As for a given diffusivity $\delta$, Stokes number $\St$ and pressure scale height $H/R$, $l_c$ scales linearly with distance $R$ to the star, the volume of the pebble cloud scales as $R^3$. And at the same time the Hill density drops as $R^{-3}$, which indicates that the mass $m_c$ included in our pebble cloud of size $l_c$ is first order independent from the distance to the star
\begin{equation}
m_{c} = \frac{4 \pi}{3} \lcrit^3 \rho_{\rm Hill} = \frac{1}{9} \left(\frac{\delta}{\St}\right)^{\frac{3}{2}} \left(\frac{H}{R}\right)^3 M_\sun,
\end{equation}
but only a function of pebble size, turbulence, aspect ratio of the disk and stellar mass. Note that there are no additional dependencies on the actual metallicity of the solar nebula (respectively protoplanetary disk), nor its mass or density profile.

When we now calculate a diameter $a_\mathrm{eq}$ for a \correctedold{solid body} of equivalent mass as this cloud of mass $m_c$, then
this is not to claim that the cloud will collapse at 100 $\%$ efficiency into precisely one planetesimal, it rather shall express what size ranges are possible in the collapse and potential subsequent fragmentation. To accommodate for this uncertainty we can incorporate the collapse efficiency $q$ which describes what fraction of the pebble cloud ends up in one individual planetesimal.

To convert mass into a size one needs a density, which makes a difference whether you form some fluffy comet with $\rhoSolid = 0.5 {\rm g}/{\rm cm}^3$ or a an asteroid with a mean density of $\rhoSolid = 2 {\rm g}/{\rm cm}^3$. For convenience we therfore use the density of our sun as density \correctedold{$\rhoSolid = 1 {\rm g}/{\rm cm}^3$}, which makes our estimate much simpler. The error in size we introduce is thus $ +- 20\%$, which is currently beyond the precision of our theory anyway. This is also consistent with the conversion of mass into equivalent size as done in the main part of \citet{Nesvorny2010}\footnote{Yet in the appendix they used $\rhoSolid = 2 {\rm g}/{\rm cm}^3$, Andrew Youdin, private communication.}.
Thus our equivalent body will have a diameter of
\begin{equation}
a_\mathrm{eq} = 2 \lcrit \left(\frac{\rho_{\rm Hill}}{\rhoSolid}\right)^{1/3}.
\end{equation}
and the possible forming planetesimals will be slightly smaller, as
$q$ enters the size only weakly:
\begin{equation}
\plntDiam = q^{1/3} a_\mathrm{eq}.
\end{equation}
Combining equations (6) and (7), and expressing $a_\mathrm{eq}$ in terms of solar radii one arrives at the relation
\begin{equation}
a_\mathrm{eq} = 
\sqrt[3]{\frac{8 \rhoSun \,\, }{9 \rhoSolid}}  \sqrt{\frac{\delta}{\St}}  \frac{\scaleheight}{R}  \Rsun \, ,
\end{equation} 
where the first term is of order unity (mean density of the Sun being similar to the mean density of planetesimals) and could be neglected for order of magnitude estimates. 

\begin{figure}
	\centering
	\begin{center}
		\includegraphics[width=1.0\linewidth]{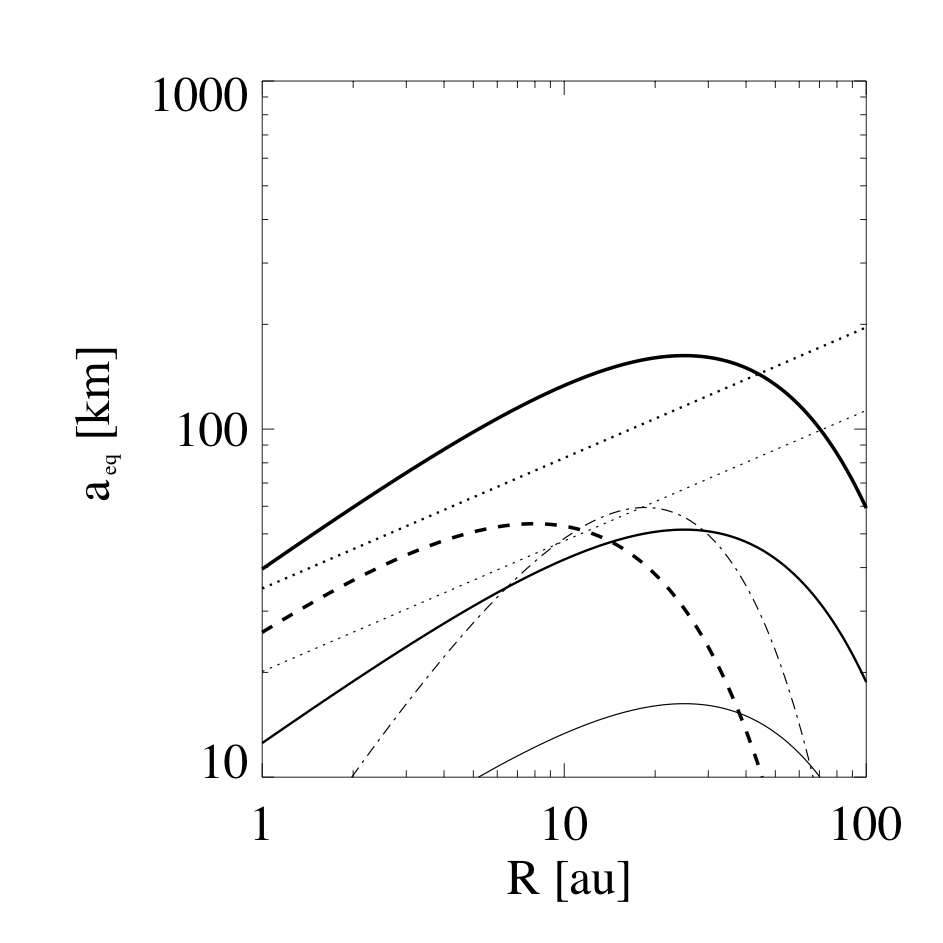}
	\end{center}
	\caption{Predicted equivalent diameter of a gravitationally unstable pebble cloud in the solar nebula model of \citet{Lenz2020} (Thick solid line) as a function of distance to the sun. The thinner solid lines are for later evolutionary stages of that gas profile, when the gas mass has decreased to $10\%$ respectively $1\%$ of the initial value. For comparison we also the equivalent diameter for the Minimum Mass Solar Nebula (MMSN) \citep{Hayashi1981} (thin dotted line) and for a $3 \times$ MMSN (thicker dotted line).
	The dashed-dotted line is also for the \citet{Lenz2020} nebula, yet for the assumption that diffusion from large scale $\alpha$ turbulence sets a lower size limit for planetesimals. 
	}
\label{fig:aeq2}
\end{figure}

\correctedold{In \citep{Lenz2020} we constrain the parameter space for the solar nebula via the influence of disk properties on planetesimal formation in our paradigm of pebble flux regulated planetesimal formation. The idea is similar to the minimum mass solar nebula \citep{Hayashi1981}, but in contrast to that model, dust does not locally grow into planetary cores, but pebbles drift large distances, before converted into planetesimals.
Therefor the nebulae in this paradigm can shallower in surface density profile of gas and still form centrally concentrated distributions of planetesimals. At the same time this shallower profile is consistent with viscous accretion disk theory for constant $\alpha$ values and also fits better observations of disks around young stars \citep{Andrews2010}.

The best parameter set to produce a planetesimal population that could explain the formation of the solar system (see \citet{Lenz2020} for details) is a relative massive yet gravitational stable disk mass of $0.1 M$, a viscosity of $\alpha = 3 \times 10^{-4}$, an exponential cut-off radius at $20$ au and for the evolution of the pebbles important a fragmentation speed of $v_{\rm frag} = 200 cm/s$. This leads to the gas density distribution as seen in Figure 2. The u-shape in the $\varepsilon_{\rm Hill}$ is direct result from the truncation radius of 20 au.

From \citet{Schreiber2018} we know that the diffusivity $\delta_0 = 2.7 \times 10^{-6}$ for $\St = 0.1$ and $\varepsilon = 10$ roughly scales inversely with $\varepsilon$ in the range of interest $10-100$ and about linear with the Stokes Number, thus we use the prescription:
\begin{equation}
    \delta = \delta_0 \frac{10}{\varepsilon_{\rm Hill}}\frac{\St}{0.1}.
    \label{eq:deltaHill}
\end{equation}
The locally dominating Stokes number we can estimate from the fragmentation limit \citep{Birnstiel2012}. The particle size is determined by global turbulence ($\alpha$), where pebbles spend most of their time, before locally concentrated to Hill density, therefore $\varepsilon$ does not affect $\St$:
\begin{equation}
    \St_{\rm frag} = \frac{1}{3} \frac{v_{\rm frag}^2}{\alpha c_s^2}.
    \label{eq:fraglimit}
\end{equation}
We plot the Stokes Numbers in Fig.\ref{fig:dust}. Thus, as we are in the range of validity for Eq.\ref{eq:deltaHill} we see that $\delta / St$ simplifies to $\delta_0 \, {10}/{\varepsilon_{\rm Hill}}$
Now we receive the length scale prediction:
\begin{equation}
    l_c \propto \sqrt{\delta_0} \sqrt{\frac{10}{\varepsilon_{\rm Hill}}} H.
\end{equation}
From this we can calculate the mass of the unstable cloud with Eq.\ (10) and from that again the equivalent diameter $a_\mathrm{eq}$ if that mass is compressed into a solid body of roughly density $\rhoSolid = 1 g/cm^3$ (eq. 13).
For instance for the MMSN with $\Sigma \propto R^{-1.5}$ and $H/R = 0.025 \cdot (R/\mathrm{au})^{1/4}$ this leads to $a_\textrm{eq} \propto R^{-3/8}$, explaining the worst case size difference between for instance $3$ and $30$ au by a factor of $2.3$.

In Fig.\ \ref{fig:aeq2} we plot the resulting equivalent diameters.
The predicted sizes are not constant with radius, yet vary surprisingly little from 80 - 140 km for a region from 3 to 50 au. We also show predicted sizes when we adopt the MMSN and a 3 times MMSN and the results are also in the 30 to 100 km range. 

The size range of around 80 km for the Asteroid belt fits nicely to the measurements by \citep{Delbo2017}. We also find sizes on the order of 100 km in the Kuiper Belt region, which opens the question how to form smaller objects.
Thus we also calculate the critical sizes for a nebula that has dropped with viscous evolution to 10$\%$ and 1$\%$ of its initial mass, and see that the eqivalent size will also shrink over time. Thus at the current location of for instance Arrokoth of 40 au, the equivalent size will shrink from 50 km down to 13 km, which would argue for a late formation of Arrokoth \citep{Stern2019} with its equivalent radius of about 20 km. Of course depending on the collapse efficiency and number of multiples that formed, the birth cloud of Arrokoth might also have had a larger mass.}

\correctedold{The presented simulations in our paper were two-dimensional. Meanwhile we tested our criterion of stability $2 \lcrit < L$ also in a limited set three-dimensional simulations \citep{KlahrSchreiber2020b} and find a conformation of our findings from the present paper.

Yet eventually one has to test our paradigm for a range of dust to gas ratios, pebble sizes and total mass of the disk \citep{Gerbig2020}, which all have an effect of the critical mass to trigger collapse.
Such a detailed study does not exist yet. Studies like \citet{Schaefer2017,Simon2017, Abod2018} and several more use in fact very large boxes and form hundreds of planetesimals to study their size distribution as function of the nebula conditions. But unfortunately diffusivity was not measured in these simulations to check for the applicability of our criterion.}
\corrected{Vertical diffusion could be measured in post processing for those existing simulations by measuring the dust scale height in the turbulent state with and without self gravity \citep{2007arXiv0708.3893J}. This method is unfortunately not possible for radial diffusion, as there is no equilibrium state with gravity balanced by diffusion. If radial and vertical diffusion would be equal, one could rely on the vertical diffusion to estimate $l_c$, yet turbulence from streaming and Kelvin Helmholtz instability is known to be rather anisotropic 
\citep{JohansenYoudin2007,Schreiber2018, Gerbig2020}. So eventually one has to repeat those simulations on the size distribution of planetesimals formed via self gravity and streaming instability \citep{Schaefer2017,Simon2017,Abod2018} and then apply a particle tracker as we describe in appendix \ref{sec:determinDiff} to determine radial diffusion. In \citep{KlahrSchreiber2020b} we present such a study on 3D streaming instability and the measurement of radial and vertical diffusion, yet on much smaller scales than in the aforementioned studies.}

\correctedold{If the pebble cloud would collapse into a single object, we could directly use the equivalent size as planetesimal size.}
This is of course not to be likely as
angular momentum conservation in the spinning and collapsing cloud can lead to fragmentation into multiple planetesimals, just like it does in star formation and also headwing may drain some material from the collapsing cloud.
Yet If only $q = 1/8$ of the collapsing cloud is converted into an individual planetesimal, its size will be smaller by just a factor of $2$ in comparison to our $q = 1$ estimate. This effect alone may be sufficient to explain a size spread among asteroids.

\subsection{The effect of global turbulence}

\correctedold{If we use global turbulence in the solar nebula instead of the locally generated turbulence by the streaming instability we can also estimate a critical size and mass scale for gravitational collapse.
First we have to consider that the global turbulence will have to cascade down from the integral or driving length scale of turbulence to the the scales of $l_c$.
Assuming a Kolmogorov cascade, as the most optimistic model, we derived the diffusivity $\delta(\lcrit)$ for the case that the global turbulence would mix proportional to the traditional $\alpha$ value, acting on the integral scale of turbulence $L = \sqrt{\alpha} H$ \citep{Schreiber2018}:
\correctedold{
\begin{equation}
    \delta(\lcrit) = \frac{\alpha^\frac{1}{3}}{1+\varepsilon_{\rm Hill}} \left(\frac{\lcrit}{H} \right)^\frac{4}{3}
\end{equation}
}
considering the mass load of material ($\varepsilon_{\rm Hill} > 1$) in the dust clump reaching Hill density \citep{2007Natur.448.1022J}. As can be seen in Fig.\ 2, the Hill density has a different slope than the gas density in the midplane and the latter will also change over time. So depending on the mass and profile of the solar nebula $\varepsilon_{\rm Hill}$ may vary between $200$ at $1$ au and $20$ at $30$ au. Same may be true for $\alpha$, which may be quite different at $1$ au vs.\ $100$ au.

If we use this local diffusivity stemming from global turbulence in the estimate for the critical length-scale Eq. (\ref{eq:collapseCrit}), we find 
\begin{equation}
    \lcrit = \frac{1}{27}\frac{\alpha^{1/2}}{\left(1+\varepsilon\right)^\frac{3}{2}} \St^{-\frac{3}{2}} H.
\end{equation}
Note the strong dependency on the Stokes number in this case, which comes from the effect that the strength of diffusion got a length scale dependence.
In the case for particle induced turbulence, when diffusivity depends on the Stokes Number, $\St$ cancelled out from the equations. Yet here we have to use an explicit $\St$ as it follows from the fragmentation limit (see Eq.\ \ref{eq:fraglimit}):
\begin{equation}
    \lcrit = \frac{1}{\sqrt{27}}\frac{\alpha^2}{\left(1+\varepsilon\right)^\frac{3}{2}} \frac{c_s^3}{v_{\rm frag}^3} H,
\end{equation}
For our $\alpha = 3 \times 10^{-4}$, that we also used in the previous part, we receive equivalent sizes that are generally smaller to the ones for the pure streaming case (See Fig.\ \ref{fig:aeq2}) and thus global turbulence does here not play a role to set the smallest scales. Yet note that the influence of global turbulence depends much stronger on $\alpha$, thus already a global alpha of $\num{1e-3}$ as used in \citet{HartlepCuzzi2020} would lead to 10 times larger sizes.
Also the $v_{\rm frag}$ has a dramatic effect in this case. That alone should rule out pure external turbulence as setting the length scales for planetesimal formation in our solar system. 
Detailed 3D simulation of this scenario are of course still missing, especially as the source and related the overall strength of turbulence in the solar nebula, i.e.\ the $\alpha$ value is heavily under debate, as is the shape and extent of the turbulent cascade \citep{2018haex.bookE.138K, Pfeil2019}.}

In the clustering model \citep{HartlepCuzzi2020} the authors use $\alpha = 0.001$ and a Stokes Number of  $\St= 0.04$ to form planetesimals at $3$ au in the solar nebula, \correctedold{indicating a large fragmentation velocity of about $600 cm/sec$.} In one of the models they assume a 10 fold minimum mass solar nebula, leading to a lower necessary concentration for collapse of $\varepsilon_{\rm Hill} = 25$ in agreement to our estimate (see Fig.\ 2). In that case we get a threshold size for the pebble cloud equivalent to a planetesimal diameter of 2300 km, which is much larger than their derived lower threshold based on ram pressure as argued for in that paper of about 10 km. But in order to have such a small ram pressure threshold they had to assume to be in a zonal flow where the headwind was reduced by a factor of 30.
If the head wind was not decreased, then they would also have received a threshold of more than 1000 km, because as can be seen in their Figure 9 the lower limit due to ram pressure (blue curve) would move upward by a factor of 30 and then the intersection with the upper limit for mass loading (red curve) would fall at a size of more than 2000 km.

So whereas it is justified to reduce the headwind in a zonal flow, the effect of turbulent diffusion will not be reduced as it is an integral part of turbulent clustering. The argument that turbulent diffusion can be neglected in comparison to ram pressure \citep{Cuzzi2008} does not hold, if one is reducing the ram pressure.

\begin{figure*}
	\begin{center}
		\includegraphics[width=0.9\linewidth]{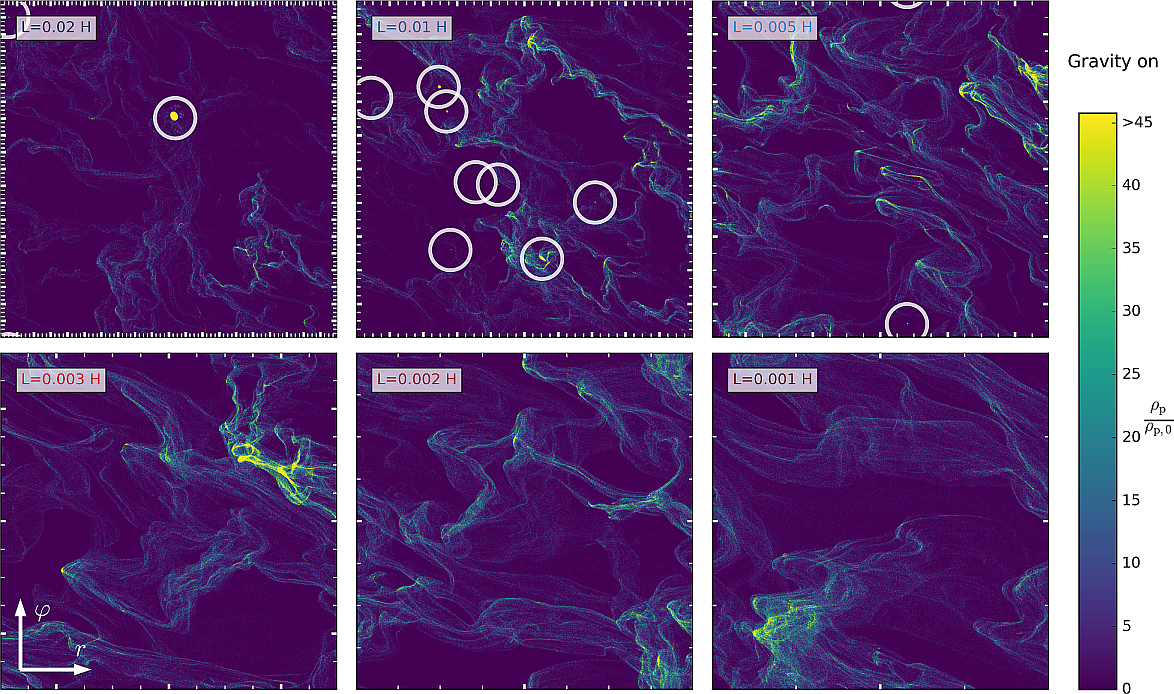}
	\end{center}
	\caption{Final particle concentration for all six \texttt{A} runs with fixed pressure gradient, see \Tab{tab:Ste-1Runs}. In white circles highlighted are all planetesimals formed, i.e. areas with a particle concentration several hundred times higher than the mean value. Only the runs with blue-colored labels (upper row) produced planetesimals as predicted from our critical length scale criteria. In these simulations the formed planetesimals are particle clumps which stay bound together after its formation. Since we set the whole simulation domain at its critical Hill density one would expect all runs to completely collapse to a single object. In contrast, our simulations show for clouds (simulation domains) much larger than its corresponding $\lcrit$ the formation of more than one object, for $L=0.01\scaleheight$ even 8 objects, and also the formation of a binary object. The runs with sizes less than its $\lcrit$ (lower row) do not collapse due to the diffusion from the underlying streaming instability. For all images the tick spacing is kept equal. This analysis is available in our online material as video covering the whole time range for all simulations up to these final snapshots.} %
	\label{fig:six_sim_grid}
\end{figure*} %

\section{Discussion}

Streaming instability in our simulations has a dual role in the process of planetesimal formation, both being contrary to each other. On larger scales the streaming instability helps to form planetesimals by concentrating dust into dense clouds and to reach Hill density, yet on small scales it prevents the formation of arbitrarily small planetesimals by diffusing collapsing clumps faster than they can collapse. 

We derived a critical pebble cloud mass to undergo gravitational collapse in the presence of turbulent diffusion, which may be either driven by streaming instability or by global gas turbulence. We showed the validity of our criterion in 2D streaming instability simulations of planetesimal formation.
\correctedold{The resulting critical pebble cloud masses for turbulence values typical for streaming instability correspond to equivalent diameters of $a_\mathrm{eq} \approx$ 100 km, and are thus compatible to from individual planetesimals of up to 100 km in diameter or several smaller ones, depending on the efficiency of the final contraction and subsequent fragmentation into multiple planetesimal systems.

Global turbulence on low levels, as needed for streaming instability in the first place, seems not to have a strong impact on small scales, beyond setting the Stokes Number $\St$. But as long as $\St$ is in a range to trigger the streaming instability, the ratio of diffusivity over Stokes Number is approximately constant as far as we know, and thus neither $\alpha$ nor the value for the fragmentation speed $v_\textrm{frag}$ have an influence onto the equivalent diameter $a_\mathrm{eq}$.

Under the assumptions that streaming instability leads to diffusion inversely proportional to the Stokes number and proportional to the dust to gas ratio as reaching Hill density, as found in \citet{Schreiber2018}, the equivalent diameter depends only on the local scale height ratio $\propto H/R$ and the inverse square root of the local dust to gas ratio of pebbles at Hill density $\propto \varepsilon_\textrm{Hill}
^{-1/2}$. As $H/R$ and $\varepsilon_\textrm{Hill}$ in a gas model of the solar nebula increase slowly with distance to the sun, the value for the eqivalent diameter $a_\mathrm{eq}$ may vary by only a factor of 2 between 3 and 30 au. Considering a steeper gas profile in the outer part of the nebula can become even smaller than at 3 au.

Global turbulence alone as setting the size of planetesimals will introduce a huge error bar, as the equivalent diameter would strongly depend on fragmentation speed $v_\textrm{frag}$ and $\alpha$, thus we can ignore this effect as long as the equivalent sizes are smaller than those set by streaming instability.}

The derived criterion supports the idea that there is an preferred initial planetesimal birth size possibly with a Gaussian distribution and not yet the power law distributions observed today. The power-law size distribution is then the outcome of planetesimal collisions and pebble accretion \citep{Johansen2015}. In fact, there is recent observational evidence that the initial size distribution of asteroids was much shallower than presumed \citep{Tsirvoulis2016} and this could be reproduced indeed by a Gaussian initial  distribution with a width of $\SI{45}{\kilo\metre}$ centered around a diameter of $80 - \SI{85}{\kilo\meter}$ \citep{Delbo2017}. The herein explained diffusion-regulated gravitational collapse of a pebble cloud is so far the only prediction for a narrow initial size-distribution of planetesimals instead of a wider power law distribution.

Planetesimals of significantly smaller size ($<$10 km) should form with a lower likelihood in this process, because turbulence can destroy initial pebble clouds of an equivalent low mass them before they collapse and thus they can only be by-products of bigger planetesimals forming. And much larger planetesimals are less frequent, because during the slowly increasing local accumulation of pebbles, the lowest possible mass will already lead to collapse.

\correctedold{We have shown that for a range of assumptions for the initial solar nebula this leads to equivalent radii of $80-140$ km, for the regions of interest,
explaining why planetesimals and thus asteroids and classical Kuiper belt objects have a kink in their distribution at about the same size. In this paradigm, larger asteroids as well as giant planet cores are the result of secondary growth processes like pebble accretion \citep{2006ApJ...639..432K,OrmelKlahr2010,Johansen2015} and collisions \citep{Kobayashi2016}, whereas smaller objects are either the outcome of a collisional fragmentation cascade \citep{Morbidelli2009} or products of the pebble cloud fragmenting in a size range of objects. These evolution processes explain the currently observed power laws above and below the initial size and thus the characteristic shape of the mass distribution of minor objects in the solar system of today is an imprint of the initial size that we explain in this work.

As the gas mass in the solar nebula decreases over time the dust to gas ratio when reaching Hill density for the pebbles will increase, which \correctedold{will} lead to \correctedold{smaller} planetesimal sizes. So in general the trend will be to first form large planetesimals and then later allow for smaller ones.}
\correctedold{Possibly at very late times with little gas left in the nebula the streaming instability was weak enough to allow for the formation of smaller planetesimals, which may be the origin of $1-10$ km sized comets. Their mass contribution should then be lower when compared to the bigger planetesimals as at late times also the pebble reservoir of the disk runs empty. Further research will have to clarify this scenario.}

But keep in mind that we derived in this paper the equivalent to the Jeans-mass in a cloud core before star formation. The gravitational collapse and the formation of multiple systems \citep{Nesvorny2010} with an initial mass function below and above the critical pebble cloud mass is a whole different story and deserves more attention.

\section{Online Content}

Movie 1: \url{https://youtu.be/gkHiluqH8HY} Simulations \texttt{Ae3L0005} and \texttt{Ae3L0005}. Both use $\St= 0.1$ particles, but only the larger box shows collapse and planetesimal formation.

Movie 2: \url{https://youtu.be/nA87-9_trUc} The evolution of $\St= 0.1$ pebbles for all 6 different box sizes in Table 1.

Movie 3: \url{https://youtu.be/CCywDPKVU8w} The evolution of $\St= 0.01$ pebbles.

\acknowledgments{We have to thank Bill Bottke for initiating this project by asking us how to explain 100 km sized objects all over the solar system. We are indebted to Anders Johansen, Marco Delbo, Allesandro Morbidelli, Jeff Cuzzi, Wladimir Lyra, Hans Baehr, Christian Lenz and Karsten Dittrich for many fruitful discussions and technical advise. Many thanks also to David Nesvorn{\`y} and Andrew Youdin for discussing their estimates of collision time vs.\ friction time with us. And last not least thanks to our anonymous referee, who suggested to include a detailed size estimate of objects in the solar nebula.
A.S.\ has been supported by the Studienstiftung des deutschen Volkes. This research was funded by  the Deutsche Forschungsgemeinschaft Schwerpunktprogramm (DFG SPP) 1385 "The first ten million years of the Solar System" under contract KL 1469/4-(1-3), by DFG SPP 1833 "Building a Habitable Earth" under contract KL 1469/13-(1-2), DFG SPP 1992: "Exoplanet Diversity" under contract KL 1469/17-1, DFG Research Unit FOR 2544 "Blue Planets around Red Stars" under contract KL 1469/15-1 and for experimental work with our colleagues in Duisburg under contract KL 1469/14-1. We also acknowledge support from the DFG via the Heidelberg Cluster of Excellence STRUCTURES in the framework of Germany'€™s Excellence Strategy (grant EXC-2181/1 - 390900948).
We received additional support by the Munich Institute for Astro- and Particle Physics (MIAPP) of the DFG cluster of excellence "Origin and Structure of the Universe" and was performed in part at KITP Santa Barbara by the National Science Foundation under Grant No. NSF PHY11-25915. 
The authors gratefully acknowledge the Gauss Centre for Supercomputing (GCS) for providing computing time for a GCS Large-Scale Project (additional time through the John von Neumann Institute for Computing (NIC)) on the GCS share of the supercomputer JUQUEEN \citep{Stephan:202326} at J\"ulich Supercomputing Centre (JSC). GCS is the alliance of the three national supercomputing centres HLRS (Universit\"at Stuttgart), JSC (Forschungszentrum J\"ulich), and LRZ (Bayerische Akademie der Wissenschaften), funded by the German Federal Ministry of Education and Research (BMBF) and the German State Ministries for Research of Baden-W\"urttemberg (MWK), Bayern (StMWFK) and Nordrhein-Westfalen (MIWF). Additional simulations were performed on the THEO and ISAAC clusters of the MPIA and the COBRA, HYDRA and DRACO clusters of the Max-Planck-Society, both hosted at the Max-Planck Computing and Data Facility in Garching (Germany). }

\vspace{5cm}
\appendix 
\section{Small or Large Clump? -- The Proper Regime for Gravitational Collapse}
\citet{Shi2013} discuss the criteria for a gravitational instability of small particles embedded in gas. They consider two cases: one in which the sound crossing time across a self-gravitating particle cloud is longer than the stopping time ($\tau_{\textrm{sound}}>\tauS$)
between particles and gas, and one case in which it is shorter. In the first case, the mixture behaves like a suspension and the clump gets stabilised
by the pressure gradient of the gas as it is getting compressed. In this case one has to consider the stability of the gas and dust mixture \citep{Cuzzi2008,Shi2013}.
In the other extreme dubbed as 'small clump' regime, when the stopping time is longer than the sound crossing time ($\tau_{\textrm{sound}}<\tauS$), one can neglect the 
effect of the gas being compressed. If we consider particles with a Stokes number of $St=0.1$  and typical dust-to-gas ratios of $\varepsilon = 3 - 100$ for collapse then 
the sound crossing distance is $\lambda = \scaleheight \frac{\St}{\sqrt{\varepsilon}} > 10^{-2} \scaleheight$, see Eq.~39 in \citet{Shi2013}. 
This distance is larger than the
clumps we consider in the main part of the paper for gravitational collapse $l_c \approx 4 \times 10^{-3} H$ and the gas can be treated as imcompressible in our considerations. 
In case of the smaller particles ($\St=0.01$) we are already entering the 'large clump' regime, which we discuss in \Sec{sec:gasPressure}, as the gas is getting slightly compressed during the
collapse of the particle cloud. Nevertheless, the collapse criteria that we derive for the 'small clump' regime still holds.
\begin{deluxetable}{lll}
	\tablecaption{Used symbols and quantities}
		\tablehead{Symbol                              & Definition                                                                            & Description                                         } 
\startdata  \\ 
		$a$                              &                $2 a = d$                                                                       & pebble radius and diameter                                               \\
		$\tauS$                             & $\tauS = \frac{a \rhoSolid \pi}{2 \Omega \Sigma_\textrm{gas}}$                                                                                       & friction time / stopping time                         \\
		$\St$                           & $\tauS\Omega$                                                                         & Stokes number                                         \\
	         $\lcrit$                            & $\lcrit = r_\mathrm{crit}$                                                          & critical length scale (radius) of a particle cloud             \\
	         $m_c$                            & $m_c = \frac{4 \pi}{3} \lcrit^3 \rhoHill$                                                          & critical mass of a particle cloud             \\
	         $a_\mathrm{eq}$                            & $a_\mathrm{eq} = 2 \lcrit \left(\frac{\rho_{\rm Hill}}{\rhoSolid}\right)^{1/3}$                                                          & equivalent contracted diameter of a particle cloud    \\
		$\mathbf{u}$, $\mathbf{v}$          &                                                                                       & gas and dust velocity                                 \\
		$v_\mathrm{frag}$          &                                                                                       & dust fragmentation velocity                                 \\
		$\Omega$                           &                                                                                       & orbital frequency                                     \\
		$R$,$z$                           &                                                                                       & semi major axis, vertical distance to midplane                                 \\
		$\cs$                               &                                                                                       & sound speed                                           \\
		$H$,$h_p$                               &                                                                                       & gas and particle scale height                                         \\
		$T_{\rm orb}$                             & $T_{\rm orb} = 2\pi/\Omega$                                                                 & orbital period                                        \\
		$\scaleheight$                      & $\scaleheight=\cs/\Omega$                                                             & gas disk scale height                                 \\
		$R$                                 &                                                                                       & heliocentric distance                                 \\
		$\Rsun$, $\rho_\mathrm{sun}$        &                                                                                       & stellar radius and solar density                      \\
		$\rhoHill$                          & $\rhoHill=9M/4\pi R^3$                                                                & Hill density                                          \\
		$\rhoCrit$                          &                                                             & shear stable could density, expressed in Hill density \\
		$\rhoSolid $              &                                                                                       & solid body density                                    \\
		$\rhoDustInit $                     &                                                                                       & initial mean dust density (simulation)                \\
		$\rhoGasInit $                      &                                                                                       & initial mean gas density (simulation)                 \\
		$\varepsilon$                          & $\varepsilon=\rhoDust/\rhoGas$                                                           & dust-to-gas density ratio                             \\
		$\varepsilon_\mathrm{max}$, $\epsInit$ &                                                                                       & maximum and initial dust-to-gas ratio (simulation)    \\
		$\varepsilon_\mathrm{Hill}$ &       $\varepsilon_\mathrm{Hill} = \rhoHill / \rhoGas$                                                                                & dust-to-gas ratio to reach Hill density   \\
		$D$,$\delta$                        & $\delta=D/\scaleheight\cs$                                                            & (dimensionless) local diffusion coefficient                 \\
		$\alpha$                        &                                                            & (dimensionless) large scale viscosity and diffusion coefficient                 \\
		$\tauFF$                            &                                                                                       & free fall time                                        \\
		$\tauC$                             &                                                                                       & contraction time (incl. friction)                     \\
		$\tauDiff$                          & $\tauDiff=r^2/D$                                                                      & diffusion time                                        \\
		$\lambda_c$                            & $\lambda_c = 2 \pi \lcrit$                                     & Jeans length scale for our simulations             \\
		$r$                                 &                                                                                       & particle cloud radius                                 \\
		$\lFree$                            &                                                                                       & mean free path                                        \\
		$L$                                 &                $[1/H]$                                                                       & simulation domain size                                \\
		$d_\mathrm{x,y}$                    &       $[1/H]$                                                                                  & simulation grid resolution                                 \\
		$\eta$                              & $\eta=\frac{1}{2}\left(\frac{\scaleheight}{R}\right)^2\frac{\de \ln \rho}{\de \ln R}$ & pressure gradient parameter                           \\
		$\Gmod$                             &                                                                                       & self-gravity parameter (simulation)                   \\ 
\enddata
\end{deluxetable}
\section{Deriving a Dust Density Criteria for Collapse: Hill Density}
Tidal forces and shear forces exerted by the host star are able to disrupt clumps if their density is less than a critical density $\rhoCrit$. For instance, a comet gets disrupted in close vicinity to a star or a planet, under the condition of differential gravity (tidal force) from the central object being stronger than the internal gravitational binding. This \textit{Roche criterion} can be expressed via a two sphere problem with radii $a/2$, mass $m$, and separation of $a$, both spheres being located at a mean distance $R$ from the central object of mass $M$. Note here, that the Roche criterion neither assumes orbital motion nor rotation, and no underlying gas flow. Thus, the maximal separation between the two spheres to overcome the tidal forces is %
\begin{equation}
	a_{\text{c}} = R \left(\frac{m}{16 M}\right)^\frac{1}{3}.
\end{equation}
This derivation directly leads to a critical breakup density for a spherical particle cloud  %
\begin{equation}
	\rho_{\mathrm{c}} = 2.5 \frac{M}{R^3},
\end{equation}
whereas the more detailed work of \citet{Chandrasekhar1967} gives a value for the Roche density of %
\begin{equation}
	\rho_{\mathrm{Roche}} = 3.5 \frac{M}{R^3}.
\end{equation}
The Roche criterion is useful to study the breakup of bodies in a close encounter, but less suited for the stability analysis of a self-gravitating particle cloud which is in an orbital motion. For this situation it is necessary to include a centrifugal potential around the primary object. %
This is the \textit{Hill criterion} in which a test particle stays bound to a secondary orbiting object, in our case it is the centre of mass of the particle cloud, with distance $a$ between test particle and center of mass.
Assuming a spherical cloud of homogeneous density distribution with a total mass of $m$ rotating at distance $R$ around a central star with mass $M$ this leads 
to an expression equivalent to the Hill sphere with radius %
\begin{equation}
	a_\mathrm{Hill}=  R \sqrt[3]{ \frac{m}{3 M}}.
\end{equation}
A test particle can only stay bound if its distance $a$ from the center of mass $m$ is $a<a_\mathrm{Hill}$.
Based on that, one can derive a critical density of that particle cloud, the Hill density: %
\begin{equation}
	\rhoHill = \frac{9}{4 \pi} \frac{M}{R^3} \approx 0.72 \frac{M}{R^3}
	\label{rhoHill}
\end{equation}
This value is smaller than the Roche density by a factor of 5, because the bound rotation of the whole particle cloud around its host star gives an additional 
stabilising effect to it. 

\correctedold{The Roche density is derived for gravity only, thus the gravitational acceleration difference by the sun $\delta g$ across a body of diameter $a$ at location $R$ scales as $\delta g = - \Omega^2 R^2 \left(\frac{1}{R_o^2} - \frac{1}{R_i^2}\right)$ with $R_{o,i} = R \pm \onehalf a$. Whereas the Hill density is calculated for an object in circular Orbit. Then the effective potential due to rotation reduces the difference in radial acceleration to $\delta g
^*= \delta g + \Omega^2 (R_o - R_i) = \delta g + \Omega^2 a$, thus a lower density is sufficient to prevent the tidal disruption of a body.}

\citet{Sekiya1983} defines his critical density for an axisymmetric 3-d annulus of particles and finds a value of %
\begin{equation}
	\rho_\mathrm{Sekiya} = 0.62 \frac{M}{R^3},
\end{equation}
in fact very close to our simple derivation.

As a conclusion of our derivation for a diffusion limited collapse, we find that only the sixth root of the particle cloud density enters the resulting planetesimal size. 
Hence, all above mentioned estimates will give nearly the same result, which is sufficient for an order of magnitude estimate.
In the following we will use the $\rhoHill$ in our calculations as critical density. In several works on planetesimal formation in fact the Hill density is used, yet it is often labelled as Roche density (e.g. \citet{Johansen2014}). So we wanted to elude a little the subtle differences in the definition.
\section{Contraction Time vs.\ Free Fall Time}
\label{sec:collapseTime}
As our gas is effectively incompressible during the collapse of the pebble cloud of the $\St=0.1$ -- $0.01$ particles, the dust-gas friction indeed alters the contraction time to longer times than the free fall time. Here, we derive a contraction timescale $\tauC$ with dependency on Stokes number $\St = \tauS \Omega$, ignoring pressure effects and assuming collapse at terminal velocity.
\subsection{Contraction time for frictional particles}
\begin{figure}
	\centering
	\begin{center}
		\includegraphics[width=0.5\linewidth]{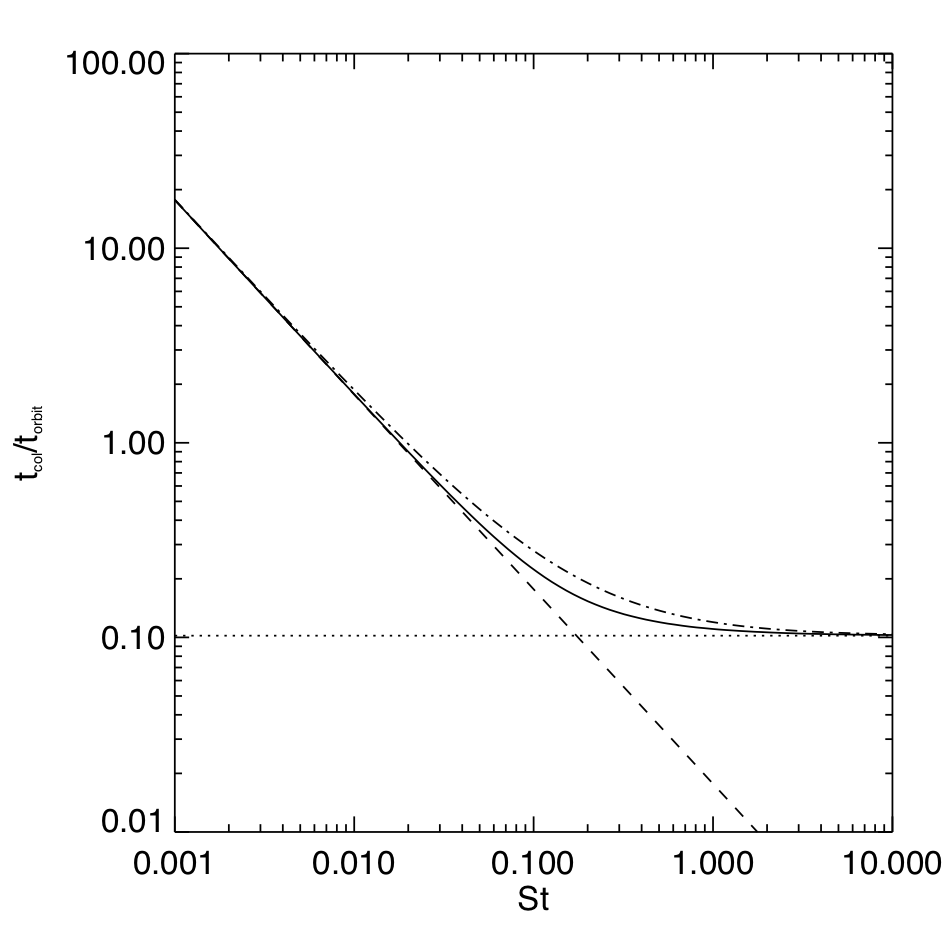}
	\end{center}
	\caption{Collapse time $\tauC$ of a dust cloud at Hill density in units of local orbital period $\torb$ as a function of the particle Stokes number. The solid line results from numerical integration of Eq.~\ref{eq:partial}. The dashed line is the analytic solution for small Stokes numbers (Eq.~\ref{eq:collapse}), the dotted line is the free fall time $\tauFF$ valid for large Stokes Numbers and the dash-dotted line is the full expression combining big and small particles, as in Eq.~\ref{eq:full}.}
	\label{f:testmode}
\end{figure}

A pressure free sphere of density $\rho$ collapses under its own gravity within a free fall timescale %
\begin{equation}
	\label{eq:freefalltime}
	\tauFF = \sqrt{\frac{3 \pi}{32 G \rho}}\, ,
\end{equation}
which for the case of Hill density $\rho = \rhoHill$ (see Eq.~\ref{rhoHill}) is a tenth of an orbital period $T_{\rm orb} = 2 \pi \Omega^{-1}$: %
\begin{equation}
	\tauFF = 0.64 \Omega^{-1} \approx 0.1 T_{\rm orb}
\end{equation}
Yet, in the case of stopping time $\tauS$ by particle-gas friction being shorter than the free fall time, particles can maximally fall at their terminal velocity \citep{Cuzzi2008}: %
\begin{equation}	
	v_\mathrm{t}(r) = -\tauS  \frac{m G}{r^2}
	\label{eq:dgl_collapse}
\end{equation}
We calculate this new frictional \textit{contraction time} from Eq.~\ref{eq:dgl_collapse} via integration: %
\begin{equation}
	\label{eq:sphereEqOfMotion}
	r(t) = \sqrt[3]{r_0^3 - 3 \tauS m G t} \quad \quad  {r\left(\tauC\right) = 0} \quad {\Rightarrow} \quad \tauC = \frac{r_0^3}{3 \tauS m G}
\end{equation}
Hence, a clump of size $r_0$ and mass density $\rho$ is expected to collapse within a collapse time %
\begin{equation}
	\label{eq:collapsetime}
	\tauC = \frac{1}{4 \pi \tauS \rho  G} .
\end{equation}
We can now express this collapse time in terms of free fall time \citep{Cuzzi2008}
and combine the both terms for long (Eq.~\ref{eq:freefalltime}) and short stopping times (Eq.~\ref{eq:collapsetime}) to %
\begin{equation}
	\tauC = \tauFF \left(1 + \frac{8 \tauFF}{3 \pi^2 \tauS}\right).
	\label{eq:full}
\end{equation}
In case that the Stokes number of the particles is smaller than the critical value
of %
\begin{equation}
	\St_\mathrm{crit} = \frac{8}{3 \pi^2 } \tauFF = 0.172
	\label{eq:stc}
\end{equation}
the simple expression %
\begin{equation}
	\tauC =  \frac{1}{9 \St} \, \Omega^{-1}
	\label{eq:simple}
\end{equation} 
is a sufficient approximation (see Fig.\ \ref{f:testmode}).

In the following section we compare our analytic estimates to a numerical integration of the 
settling process.

\subsection{Numerical test of contraction time and analytic fit}
The differential equation governing the settling process is
\begin{equation}
	\partial_t v =   - \frac{m G}{r^2} - \frac{v}{\tauS}.
	\label{eq:partial}
\end{equation}
We time integrate this equation with a Leap Frog algorithm with the initial condition $r(t=0) = r_0$ and $v(t=0)=0$. The parameters $G$ and $m$ are chosen in a way to initialize the cloud at Hill density, i.e. spreading the mass $m$ evenly over the volume $V=\frac{4}{3}\pi r_0^3$. We performed a set of simulations for different single particle sizes ranging from $\St= 10^{-3}$ to $\St= 10$, see Fig. 1. We find that the simple fit from Eq.~\ref{eq:full} is perfectly suited for all particle sizes, and even Eq.~\ref{eq:simple} gives good results up to St = 0.1.

\section{Detailed Critical Length Scale Derivation and Resulting Planetesimal Size}
Planetesimal formation happens in a shearing environment, hence one needs to ensure tidal disruption of a particle cloud as primary condition for planetesimal formation via gravitational collapse. 
As in the main paper we start from setting diffusion time and collapse time equal. Here, lets consider small particles with $\St \ll 1$, falling at terminal velocity:
\begin{eqnarray}
\tauC &=& \tauDiff\\
\frac{\rC^3}{3\tauS m G} = \frac{1}{4\pi\tauS \rhoInt G} &=& \frac{\rC^2}{D} = \frac{\rC^2}{\delta H \cs} \nonumber
\end{eqnarray}
Where, $m=4/3 \pi \rC^3 \rhoInt$ is the bulk mass of the cloud with radius $\rC$. The diffusion timescale stems from Fick's second law of diffusion, $\dot{\rho} = D \nabla^2 \rho $. We simplify, by expressing the internal cloud density in terms of Hill density via a scaling parameter $f$:
\begin{equation}
\rhoInt = f\cdot\rhoHill = \frac{9f}{4\pi}\frac{M}{R^3}
\label{eq:rhoInt}
\end{equation}
With this simplification, the critical cloud diameter is
\begin{equation}
\label{eq:collapseCrita}
\rC = \lC = \frac{1}{3} \sqrt{\frac{\delta}{f\cdot\St} H}\, .
\end{equation}
This expression is valid for all $f$ as long as the condition for shear and tidal stability is given. 
\subsection{Jeans Length for planetesimal formation}
A full analysis of the stability of dust under self-gravity embedded in gas would lead to a Toomre analysis. There one performs a linear analysis of the problem and derive a stability criterion from a
dispersion relation, following a mixed case of \citet{GoldreichWard1973} and \citet{Safronov1969}, similar yet not identical to the secular gravitational instability \citep{Ward2000}. 
The resulting Toomre criterion would tell us whether there was a fastest growing mode, which as we would see is larger than our simulation domain. Toomre combines two obstacles for gravitational collaps: A: tidal forces, i.e. angular momentum conservation on large scales, and B: thermal pressure on small scales. 

Thus we will focus on the small scales, which is just the Jeans length, and we will see that even the Jeans length is larger than our boxsize.

The difference in the derivation here is that instead of the thermal pressure, we us the diffusion flux of particles $j = -D \nabla \rho$ to be in equilibrium with sedimentation.
Starting with the continuity Eq. for the dust particles:
\begin{equation}
\partial_t \rho + \nabla v \rho =  0
\end{equation}
The flux $\rho v$ is given by diffusion and sedimentation under self gravity with potential $\Phi$,
\begin{equation}
\partial_t \rho - \nabla \left( D \nabla \rho + \tau \rho \nabla \Phi\right) =  0,
\label{eqcon}
\end{equation}
where we use the terminal velocity ansatz $v = \tau g = - \tau g \nabla \Phi$, with $g$ the gravitational acceleration.
With a linearisation in density $\rho = \rho_0 + \rho'$, which will also lead to a linearisation in $\Phi$, 
we can simplify this using $\nabla \rho_0 = 0$ and $\nabla \Phi_0 = 0$ to 
\begin{equation}
\partial_t \rho' - D \nabla^2 \rho' - \tau \rho_0 \nabla^2 \Phi' =  0.
\end{equation}
We replace $\Phi'$ via the Poisson equation 
\begin{equation}
\partial_t \rho' - D \nabla^2 \rho' - \tau \rho_0 4 \pi G \rho' =  0.
\end{equation}
With the usual plane wave ansatz $\rho' = \rho_a e^{-i\left(\omega t - k x \right)}$ we get to 
\begin{equation}
-i \omega  + k^2 D - \tau \rho_0 4 \pi G=  0,
\end{equation}
and it is obvious that all waves with $k < k_c = \sqrt{4 \pi G \rho_0 \frac{\tau}{D}}$ will be unstable and collapse.
If we put in the Hill density we receive:
\begin{equation}
k_c = 3 \sqrt{\frac{\tau}{D}}\Omega,
\end{equation}
which again we express in wavelength:
\begin{equation}
\lambda_c = 2 \pi \frac{1}{3} \sqrt{\frac{\delta}{\St}}H = 2 \pi l_c,
\end{equation}
thus our numerical setup is linear stable to self gravity, because $L < 2 \pi l_c$, but once streaming instability has created non-linear perturbations, those can collapse to planetesimals, if diffusion is weak enough as stated by $L > 2 l_c$.

\subsection{Particle scale height}
We can also ask for the scale height that a particle layer would have if being in equilibrium 
between sedimentation and turbulent diffusion. For no selfgravity and particles with stokes numbers larger than the dimensionless diffusivity $\delta$ this would be the well known result 
\begin{equation}
h = \sqrt{\frac{\delta}{\St}}H,
\end{equation}
that is if vertical gravity stems purely from the star.
But in case of reaching Hill density in the mid-plane, self gravity is an order of magnitude stronger than 
the stellar gravity. We reuse above condition for equilibrium from Eq. \ref{eqcon}:
\begin{equation}
\partial_t \rho = - \partial_z \left( D \partial_z \rho + \tau \rho \partial_z \Phi\right) =  0,
\end{equation}

\begin{figure}
	\centering
	\begin{center}
		\includegraphics[width=0.5\linewidth]{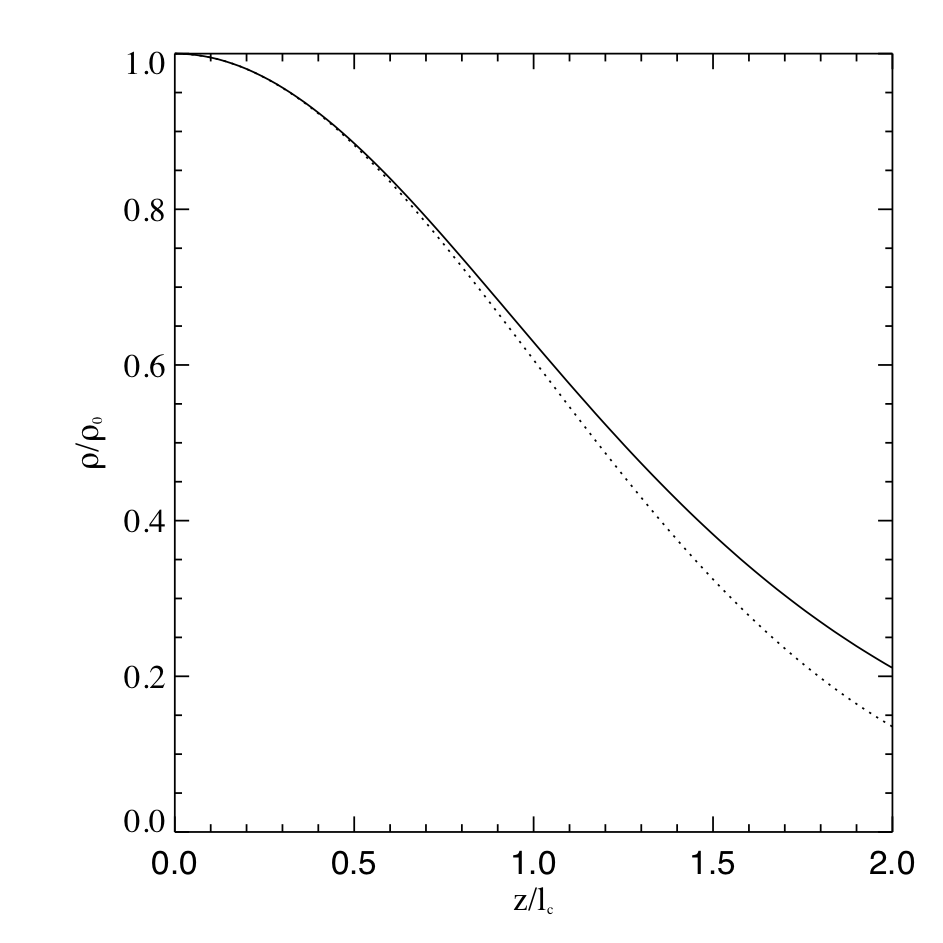}
	\end{center}
	\caption{Vertical dust distribution for particles in equilibrium between self gravity at Hill density in the midplane and vertical diffusion. The solid line is the correct solution, see Eq.\ \ref{eq:collapseCrit2}, and the dashed line is a Gaussian with the same characteristic length $\lcrit$ for comparison.
	\label{f:layer}}
\end{figure}

Which leads to the differential eq.:
\begin{equation}
\partial_z^2  \ln \rho = - \frac{\tau}{D} 4 \pi G \rho.
\end{equation}
If we express $\rho$ in Hill density as above, and combine the remaining terms in our 
above defined critical length $\lcrit$, this is simply:
\begin{equation}
\partial_z^2  \ln \rho = - \frac{1}{\lcrit^2} \frac{\rho}{\rhoHill},
\end{equation}
which has the analytic solution 
\begin{equation}
\rho(z) = \rhoHill \left[1 - \tanh^2\left(-\frac{z}{\sqrt{2} l_c} \right)\right] = \frac{\rhoHill}{\cosh^2\left(-\frac{z}{\sqrt{2} l_c} \right)} .
\end{equation}
Thus $l_c$ is a truly versatile value for dust layers and clumps likewise (see Fig.\ \ref{f:layer}).
Note that collapse here can only occur in 2D or 3D, because then the collapse time shrinks faster with length than
diffusion time can increase. But in 1D a flat layer would re-expand if compressed below its vertical equilibrium height $l_c$ because the gravitational potential at the surface of a flat sheet does not depend on the thickness of that sheet. 
\section{Numerical Test on the Critical Length Scale Criteria}
\subsection{Used method: \pc}
For our numerical investigations we use the \pc\footnote{\url{http://pencil-code.nordita.org/}} (see \citet{YoudinJohansen2007}, \citet{Brandenburg2001}, \citet{Brandenburg2002}, \citet{Brandenburg2005} for details). The \pc \, is a general numerical solver, here used on a finite-difference hydrodynamical code using sixth-order symmetric spatial derivatives and a third-order Runge-Kutta time integration. The simulations are done in the shearing-sheet approximation (see \citet{Goldreich1965},\citet{Hawley1992}, \citet{Brandenburg1995}), a Cartesian coordinate system co-rotating with Keplerian frequency $\Omega$ at distance $R$ from the star. Thus, all quantities have to be interpreted as being local, e.g. the shear is linearized via
\begin{equation}
	\label{eq:linear_shear}
	u_y^{(0)}=-(3/2)\Omega x,
\end{equation}
with $x$ the radial coordinate in the simulation frame. All simulations are dimension free hence time and scaling can be chosen arbitrary, e.g. by defining the distance to the star. The coordinate system $\left(\vec{e}_x, \vec{e}_y, \vec{e}_z\right)$ can be identified as $\left(\vec{e}_r, \vec{e}_\varphi, \vec{e}_z\right)$. The boundary conditions are periodic in $y$-/$z$- and shear-periodic in $x$-direction. We perform our simulations in a 3-d \citep{2007Natur.448.1022J}, but with only one grid cell in z-direction, see \Sec{sec:collapse2dcylinder}. This means we one suppress modes in vertical direction, diffusion by SI is weaker in this direction anyway, and also avoid the necessary to use a thin disk approximation or even a 2-d gravity approach. To ensure dissipation on grid scale, sixth-order hyperdissipation terms are used \citep{Lyra2008, Lyra2009}, since the \pc high-order scheme has only marginally numerical dissipation.

All particles used in the simulations are Lagrangian super-particles each representing a swarm of identical particles interacting with the gas as a bulk. Their properties, e.g. density, is smoothed out to the neighboring grid cells via the Triangular Shaped Cloud (TSC) scheme \citep{YoudinJohansen2007}.

\subsection{Physical model}
The simulations solve the Navier-Stokes equation for the gas and the particle motion in a shearing box approximation. The gas velocity $\vec{u}$ relative to the Keplerian shear is evolved via %
\begin{eqnarray}
	\frac{\p \vec{u}}{\p t} + (\vec{u}\cdot\nabla)\vec{u}+u_y^{(0)}\frac{\p \vec{u}}{\p y} &=& \; 2\Omega u_y \hat{x}-\frac{1}{2}\Omega u_x \hat{y} +\Omega^2z\hat{z}\\ -\frac{1}{\rhoGas} c_s^2\vec{\nabla} \rhoGas 
	&-& \frac{\rhoDust / \rhoGas}{\tauS}\left[\vec{u}-\vec{v}\right] + f_\nu(\vec{u},\rhoGas),
\end{eqnarray}
with second and third terms on the left-hand side the advection terms by the perturbed velocity and by the shear flow, respectively. On the right are the terms for Coriolis force, the pressure gradient (with $\vec{P}=c_s^2\vec{\nabla} \rho$), the particle-gas drag interface and the viscosity term. The pressure gradient is split up into a global enforced pressure gradient via $\eta$, see Tab. 1, that is acting on the gas rather than the particles (compare with Athena code) and in the local contribution from actual evaluated gas density in the simulation domain.

The gas density is evolved with the continuity equation %
\begin{equation}
	\frac{\p \rhoGas}{\p t} + \left(\vec{u}\cdot\vec{\nabla}\right)\rhoGas + u_y^{\left(0\right)} \frac{\p \rhoGas}{\p y}= -\rhoGas \vec{\nabla}\cdot\vec{u} + f_\mathrm{D}\left(\rhoGas\right).
\end{equation}
The particles are evolved via %
\begin{equation}
	\frac{\D x^{\left(i\right)}}{\D t} = \vec{v}^{\left(i\right)} + v_y^{\left(0\right)}\hat{y},
\end{equation}
with Keplerian orbital velocity $v_y^{\left(0\right)}$ and particle velocity $\vec{v}^{\left(i\right)}$, which is evolved similarly to the gas
\begin{equation}
	\frac{\D \vec{v}^{(i)}}{\D t} =  \; 2\Omega v_y^{(i)} \hat{x}-\frac{1}{2}\Omega v_x^{(i)} \hat{y} - \Omega^2 z \hat{z} + \frac{1}{\tauS} [\vec{v}^{(i)}-\vec{u}(x^{(i)})]
\end{equation}
but without the gas pressure gradient acting on it. The interface between gas and particles is determined by the gas and dust densities $\rhoGas$ and $\rhoDust$ and friction time $\tauS$. As it is typically, in this paper the friction time is expressed in orbital periods, called the Stokes number $\St=\tauS\Omega$.

The gravitational potential is calculated by solving the non-dimensinal form of the Poisson equation
\begin{equation}
	\left(\scaleheight\nabla\right)^2 \Phi/\cs^2=\Gmod\frac{\rhoDust}{\rhoGas}
\end{equation}
via the Fourier method \citep{2007Natur.448.1022J}, hence $\Phi\left(\vec{x}\right)=\sum
_k \Phi_k \exp\left(\ii \vec{k}\cdot\vec{x}\right)$ with spatial wave number $\vec{k}$ and $\Phi_k=-4\pi \Gmod\tilde{\rho}_k/\left|\vec{k}\right|^2$. Here, $\tilde{\rho}_k$ is the Fourier amplitude and $\Gmod$ is the self-gravity parameter which one gets by adopting the \pc unit system for shearing box simulations of $c_\mathrm{s}, \gamma, \rhoGasInit, \Omega, \scaleheight = 1$.

\subsection{Numerical model}
\label{sec:numericalModel}
First and foremost we are interested in the particle diffusivity $\delta$ since it will allow us to predict whether collapse can occur or not. Therefore, we start with gravity switched off in order to get the simulation in a saturated streaming instability state. In this gravity free state a particle tracking scheme can be used to measure the pure diffusivity of the streaming instability, as explained in section~\ref{sec:determinDiff}. Once this is achieved, gravity is switched on in a fashion that sets the initial dust density to be Hill density, that is sufficient to ensure collapse if streaming instability does not prevent it.

Since we demand a certain dust-to-gas ratio for our study, the only way to set the initial simulation dust density to Hill density is by altering the gravitational constant $\Gmod$ such that 
\begin{equation}
	f\cdot\rhoHill =  \frac{9f}{4\pi}\frac{M}{D^3} {=} \rhoTotal = \rhoDust,
\end{equation}
with stellar mass $M$ and distance of the particle cloud from the central star $D$. We set the total density to the particle density, since the gas density stays constant throughout the collapse and thus does not contribute to the gravity acting in the simulation. Here, the parameter $f$ is introduced to alter the internal density in terms of Hill density. This equation can be simplified by using the dust-to-gas ratio $\eps = \frac{\rhoDust}{\rhoGas}$ and by using $M=\frac{D^3 \Omega^2}{\Gmod}$, resulting in
\begin{equation}
\frac{9f}{4\pi} \frac{\Omega^2}{\hat{G}} = \rhoGas\cdot\eps.
\end{equation}
Solving for $\hat{G}$ results in
\begin{equation}
	\Gmod = \frac{9f}{4\pi}\frac{\Omega^2}{\rhoGas\cdot\eps}.
	\label{eq:gravConst}
\end{equation}
In our case with $\epsInit=3$, $\Omega = 1$, $\rhoGasInit=1$ and $f=1$ we have to set the gravitational constant to $\Gmod=0.2387$.

\subsection{Model setup}

For our case study each run uses 16 CPUs in x- and 8 CPUs in y- and 1 CPU in z-direction, to evaluate a 256$\times$256$\times$1 grid cells simulation domain and the particles therein. The runs are initiated with 10 particles per grid cell, thus having $655,360$ particles per run. Two types of single-species particles are used: The \texttt{A} runs have $\St=0.1$ particles and the \texttt{B} runs have $\St=0.01$, see \Tab{tab:simResults}. Particles start randomly distributed but match an initial average density of $\epsInit=3$ and are initiated in gas-dust drag force equilibrium \citep{Nakagawa1986}. Since the simulation domain is representing a dust particle cloud of a certain size, we vary this size around $\lcrit$, see \Tab{tab:simResults}.

All simulations start with gravity switched off to ensure streaming instability being saturated before collapse is allowed. This is done by activating gravity after $t=1.59$ orbits (\texttt{A} runs) or $t=4.77$ orbits (\texttt{B} runs), with gravitational constant set as derived in Eq.~\ref{eq:gravConst}, i.e. setting the initial density to the critical Hill density.

Additionally to the main runs, we study the impact of variation in pressure gradient $\eta$ on our criteria for the case of $\St=0.1$ particles. Hence, we set up additional simulations of the two simulations around $\lcrit\approx \onehalf L$ with $2\cdot\eta$ and $0.5\cdot\eta$, see \Tab{tab:Ste-1Runs}.
\subsection{Collapse criteria validity in our 2-d simulations}
\label{sec:collapse2dcylinder}
Since full 3-d simulations are highly expensive compared to 2-d, we here use a setup were the z-dimension has a single grid cell. Consequently, we use the same gravitational force as in a full 3-d setup, but instead of evaluating the collapse of a 3-d sphere we evaluate the collapse of an infinitely extended 3-d cylinder. Nevertheless, here we show that the free-fall and collapse times are in fact identical in both cases.

The gravitational force on the surface of a 3-d sphere with radius $R$ and mass $m = 4/3 \pi \rhoInt R^3$ is
\begin{equation}
F_\mathrm{g,sph} = -\frac{GmM}{R^2}\hat{r},
\end{equation}
with unit vector $\hat{r}$, since, we can collapse the whole cylinder to a single line of mass M. The gravitational force on the surface of a 3-d cylinder, around this line of mass with linear density $\lambda = M/L$, with cylinder length $L$, one gets by calculating the gravitational potential $\Delta\Phi=4\pi G\rhoInt$:
\begin{equation}
F_\mathrm{g,cyl} = -\frac{2 Gm \lambda}{R}\hat{r}
\end{equation}
Following section~\ref{sec:collapseTime}, we get the equation of motion for a particle on the cylinder surface as it collapses as
\begin{equation}
r\left(t\right) = \sqrt{r_0^2 - 4\pi\tauS \rhoInt G t}.
\end{equation}
Already at this point one can see a clear parallel to Eq.~\ref{eq:sphereEqOfMotion}, since they only differ in the exponent of the root function. This dependence then eliminates when solving for collapse time $\tau_\mathrm{c}$ and both, for spheres and cylinders, the collapse time is
\begin{equation}
\tau_\mathrm{c, cyl}   = \frac{1}{4\pi \tauS \rhoInt G} = \tau_\mathrm{c, sph}.
\end{equation}
We want to stress out, that the point of our 2-d model is rather to show that our analytic criterion of balancing the particle cloud contraction with diffusion is properly predicting the outcome of this non-linear simulations. The fact that contraction time is identical for 2-d and 3-d configurations explains why the criterion is also suited for our 2-d simulations.
\begin{figure*}
\gridline{
          \fig{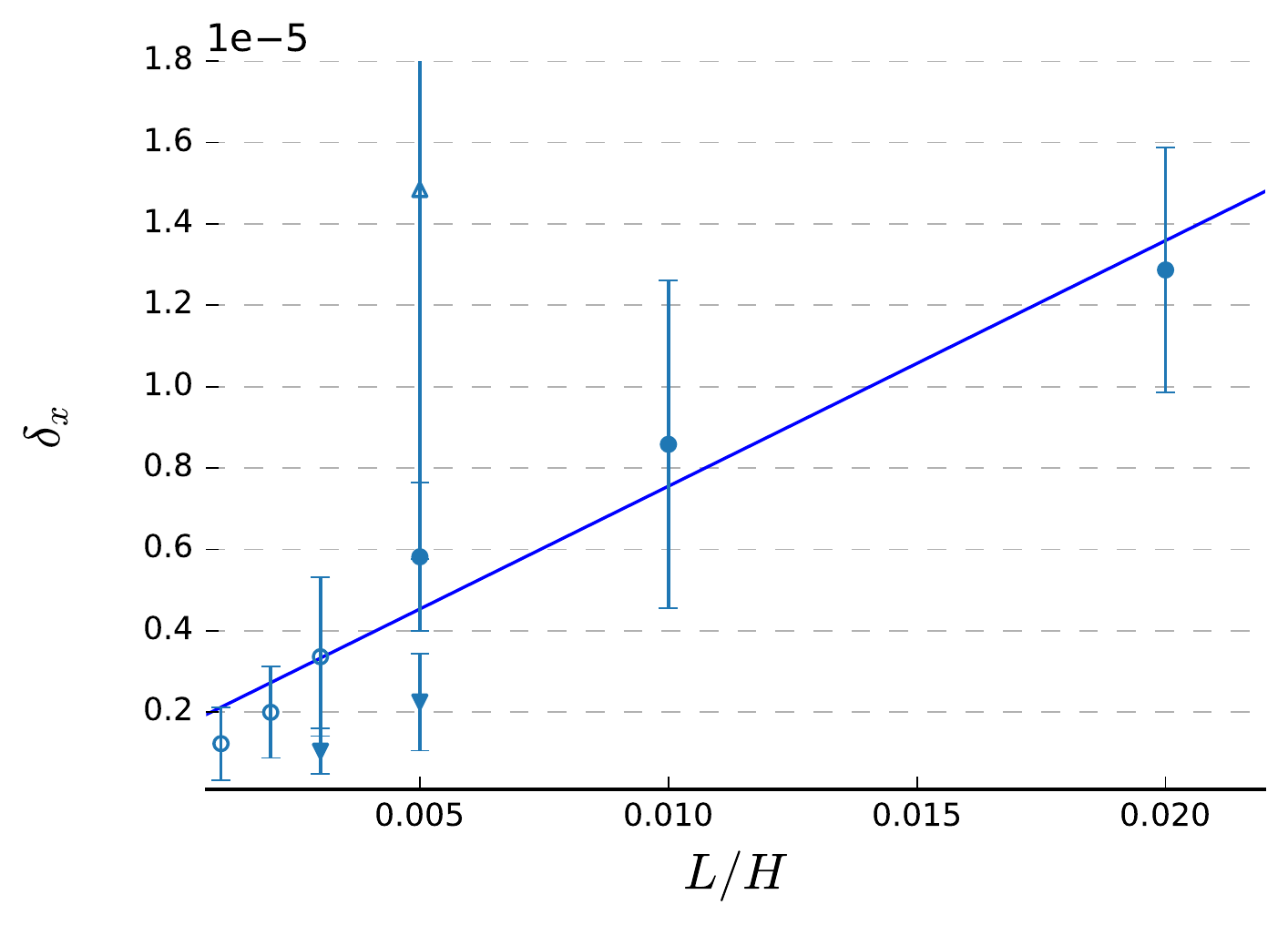}{0.45\textwidth}{(a) \texttt{A} 
          runs with $\St=0.1$
          }
          \fig{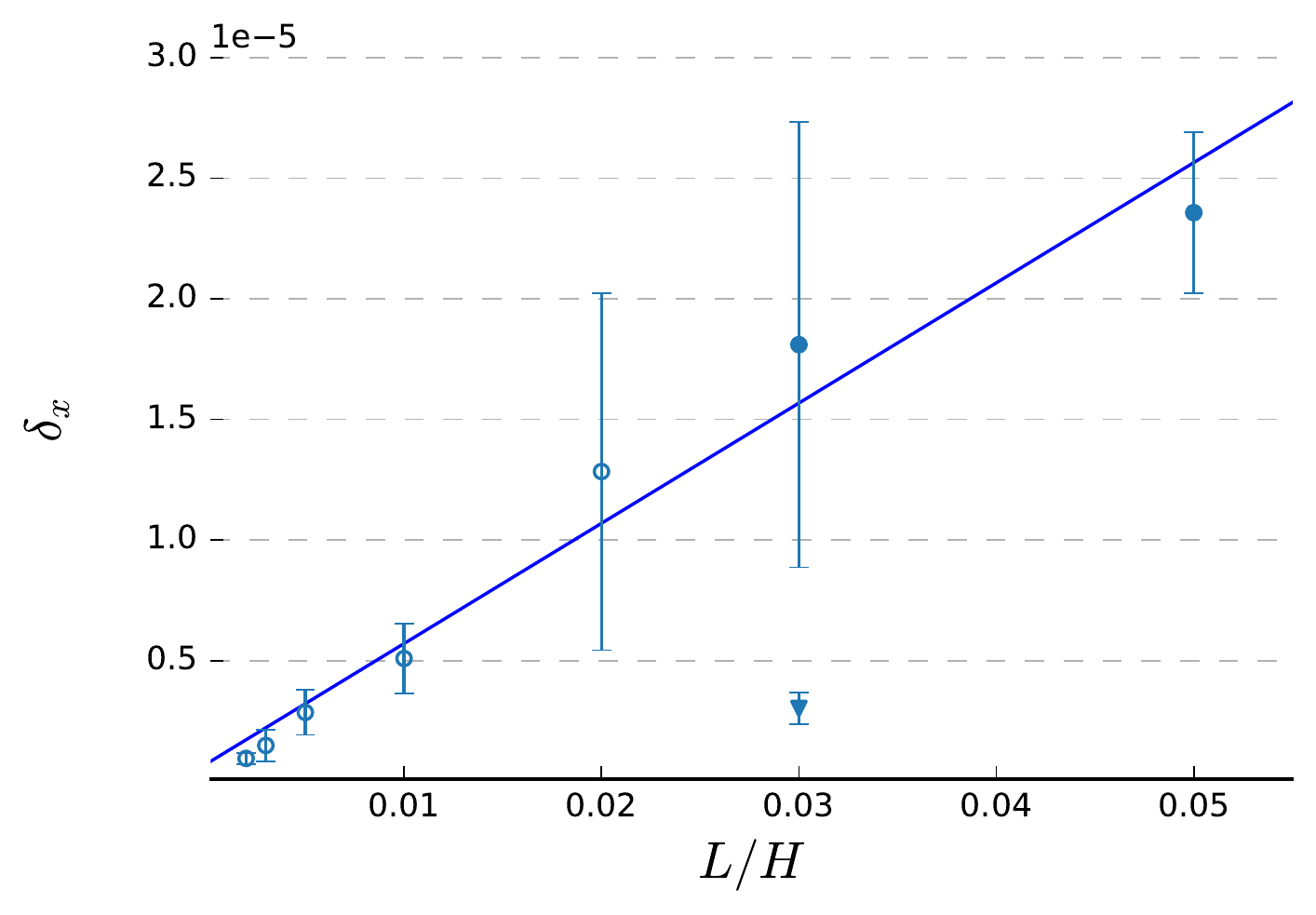}{0.45\textwidth}{(b) \texttt{B} runs with $\St=0.01$
          }
          }
          \caption{Radial diffusion over simulation domain size. Circles indicate that the standard pressure gradient and dust to gas ratio was applied. 
          If a symbol is filled it marks a run in which collapse occurred.
          The triangle pointing up in the \texttt{A} run indicates the run with double the pressure gradient (hp), triangles pointing down indicate runs with half the pressure gradient (lp). The triangle pointing down in the \texttt{B} run uses twice the the dust-to-gas ratio.
          Straight lines are fits to the standard simulations, i.e.\ circles. Slopes of this fit are $p_\mathtt{A}=\num{6.04e-4}$ and $p_\mathtt{B}=\num{4.98e-4}$.
          \label{fig:diffusion}}
%
\end{figure*} %
\subsection{Measuring particle diffusivity in a shear flow}
\label{sec:determinDiff}

The critical quantity preventing collapse is diffusivity $D$ of the streaming instability, which can be expressed in disk units of orbits $\Omega$ and sound speed $c_\mathrm{s}$
\begin{equation}
\delta = \frac{D}{c_\mathrm{s}^2/\Omega}.
\label{eq:diffusioninorbs}
\end{equation}
The diffusion is measured by tracking the position of a sample of at least $10^4$ super-particles and recording their travel distance with time. The time derivative of the variance of the resulting travel distance histogram gives directly the diffusion $D$ by using 
\begin{equation}
\label{eq:diffusion}
D=\frac{1}{2}\frac{\p \sigma_\mathrm{Gauss}^2}{\p t},
\end{equation}
with Gaussian variance $\sigma_\mathrm{Gauss}^2$ of the distribution, as introduced in \citet{JohansenYoudin2007}. This leads to a mean travel distance from the initial particle positions of $\left<r^2\left({t}\right)\right>=D{t}$ after a time ${t}$. The diffusivity is measured in the  saturated phase of the streaming instability for each simulation before gravity is switched on.

\begin{figure}
\gridline{
          \fig{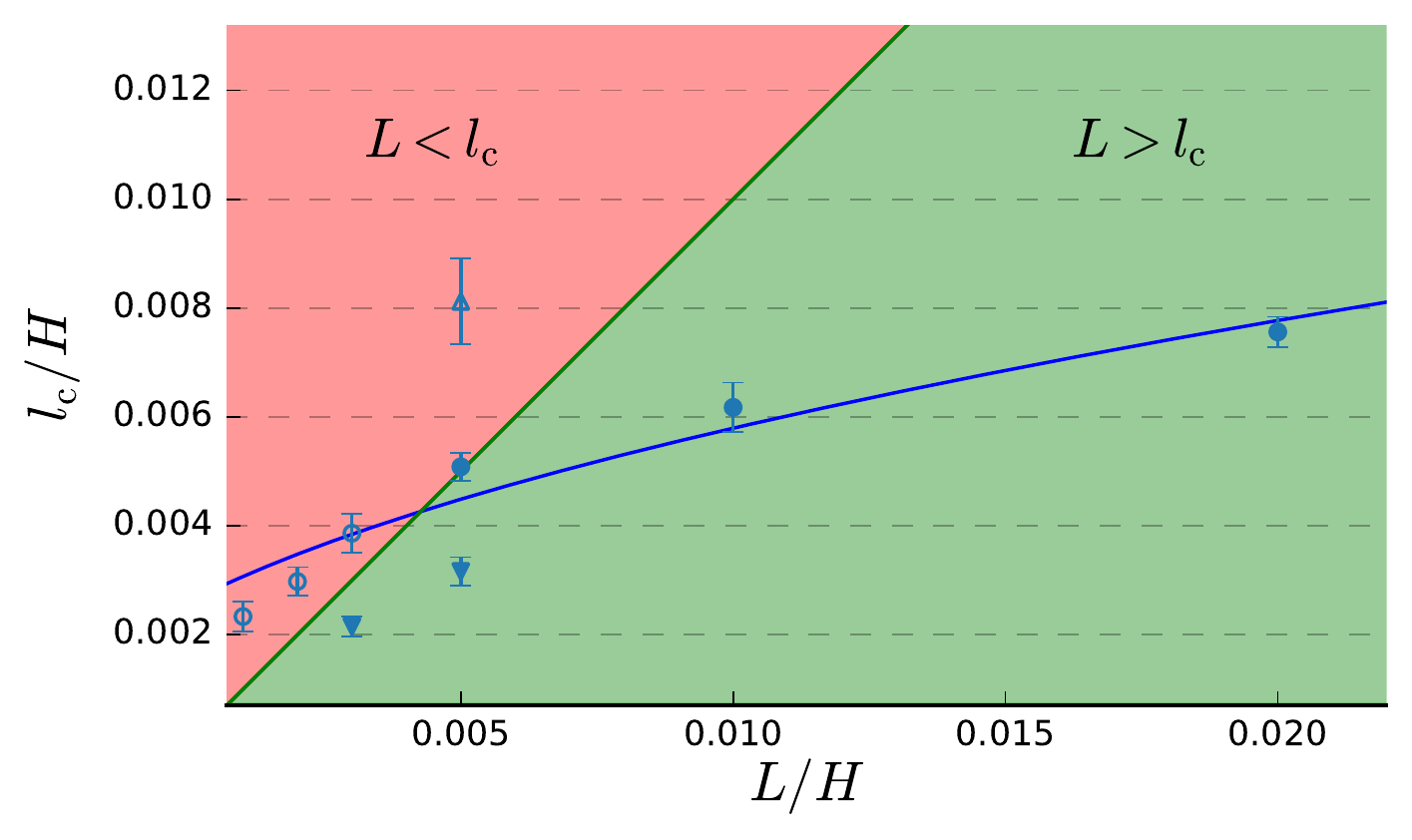}{0.5\textwidth}{(a): Additional \textbf{A} runs with \textbf{$\St=0.1$}}
          \fig{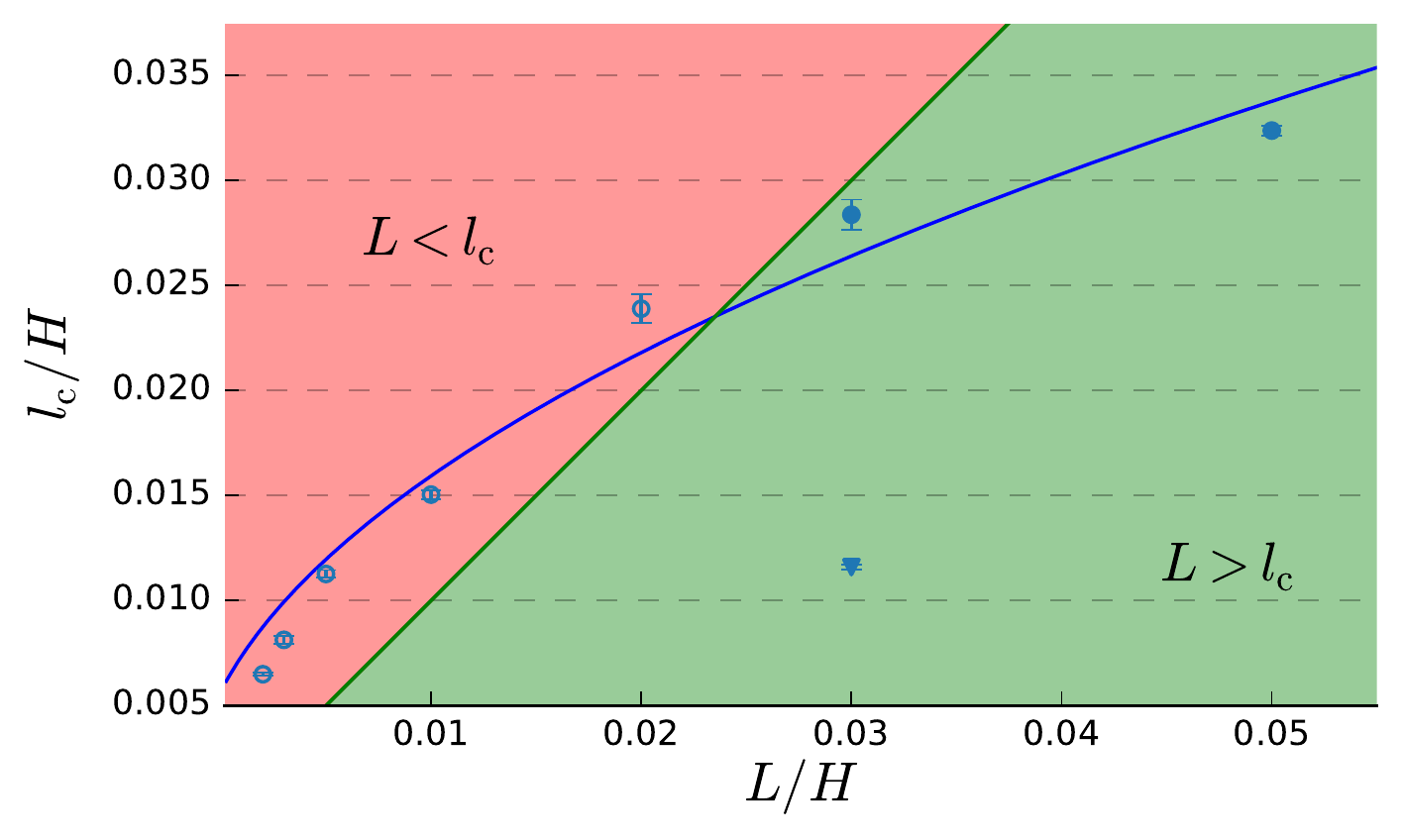}{0.5\textwidth}{(b): Additional \textbf{B} runs with \textbf{$\St=0.01$}}
          }

\caption{\textbf{Numerical results compared with analytic prediction} With domain size $L$ on the x-axis we plot the critical length scale $\lcrit$. This scale is determined by measuring the diffusivity of the pure streaming instability before switching on self-gravity. The red region indicates $L < \lcrit$ where no collapse should be possible whereas in the green region $L > \lcrit$ collapse should occur. We find agreement between our prediction and the simulation results: All simulations with filled symbols did collapse and the ones with open symbols did not.}
\label{fig:lcritA}
\end{figure}

\subsection{Error bar estimation}
The error in diffusivity $\Delta\delta$ is estimated by calculating the standard deviation of diffusivity time series $D\left(t\right)$, see Eq.~\ref{eq:diffusion}. From this one gets the error in the critical length scale via
\begin{equation}
\Delta \lcrit = \frac{2}{6} \Delta\delta \cdot \delta^{-\frac{1}{2}}
\end{equation}
\subsection{Increasing gas pressure during the collapse}
\label{sec:gasPressure}
Gas pressure might increase within the collapse phase due to friction of particles acting on the gas, dragging it along while collapsing. The reason is that the collapse phase is a situation of high dust concentration, meaning momentum of the dust is large, and Stokes number is low, so its motion is well coupled onto the gas. \citet{Shariff2015} describe this effect in numerical 1-d models. They claim it can lead to oscillations in internal dust density and particle cloud core size, hence delaying the collapse for a certain parameter range, i.e. initial dust-to-gas ratio of $\eps=10$ to $100$.

\begin{figure}
	\begin{center}
		\includegraphics[width=\linewidth]{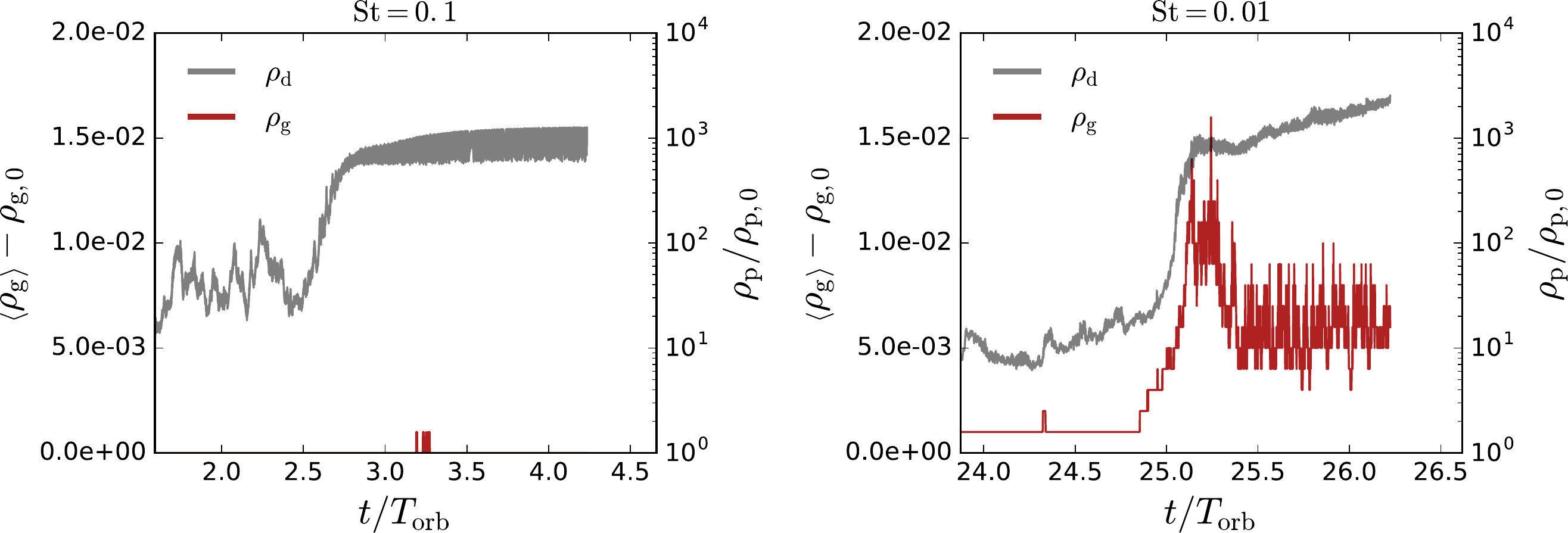}
	\end{center}
	\caption{Change in gas pressure (red) withing the collapse phase of the dust cloud (grey) for both investigated Stokes numbers. Only last orbit before collapse is shown. We only find a small change in gas density that correlates with the collapse phase for $\St=0.01$, but no hints on a strongly delayed collapse by oscillations. They may remain unresolved.} %
	\label{fig:gasPressure}
\end{figure}

In our simulations we also check for changes in gas pressure. Since we perform our simulation in the ideal gas limit, we have, since $P=\rhoGas \cs^2$, to check our simulations for an increase in gas density that correlates with particle cloud collapse. Figure \ref{fig:gasPressure} shows the time-series of gas and dust density for the critical collapsing cases for both investigates Stokes numbers: \texttt{Ae3L0005} and \texttt{Be3L003}. We find for $\St=0.1$ no change in gas density. The strongest change in gas density is happening far after the planetesimal has formed. For $\St=0.01$ we indeed find a correlated increasing gas pressure, being slowly build up while the dust cloud is collapsing. But the change in pressure is with $\Delta p \approx 0.01$ rather small, consequently for this setup we can assume to not be in the suspension regime, though gas pressure might have an influence at the unresolved scales it will not prevent the collapse.
\subsection{Effects of particle collisions during the collapse}
\label{sec:collisions}
To justify that we are allowed to neglect particle-particle collisions in our numerical experiments, we have to estimate the collision timescale and compare it to the collapse timescale.
The collision time per particle is given by 
\begin{equation}
	\tau_\mathrm{coll} = \lFree / v_\mathrm{rms},
\end{equation}
i.e.\ the ratio of free mean path of a particle and the particle bulk rms-velocity. The mean free path is a function of particle number density $n$ with a certain size $a$ and their combined cross section $4 \pi a^2$:
\begin{equation}
	\lFree = \frac{1}{4 \pi n a^2}
\end{equation}
The number density has to be calculated from the particle its mass density which we express as multiples $f$ of the Hill density $\rho_\mathrm{dust} = f  \rhoHill$
\begin{equation}
	n = \frac{f  \rhoHill}{\frac{4}{3} \pi a^3 \rho_0},
\end{equation}
thus
\begin{equation}
	\lFree = \frac{a }{3 f}\frac{\rho_0}{\rhoHill},
\end{equation}
which could be explicitly calculated if we knew the actual particle size. This is only possible if one would define all physical parameters entering the relation between Stokes number and particle size, i.e. stellar mass, distance to the star, density and gas temperature and the porosity of the dust. But, we know the particle its mass density equals the Hill density or multiples of it, plus their Stokes number is to be $\St= 0.1$. With 
\begin{equation}
	\St = \tauS \Omega = \frac{a \rho_0 \Omega}{\rho \cs}
\end{equation}
this gives a size of 
\begin{equation}
	a = \frac{\scaleheight \St \rhoHill}{\rho_0 \varepsilon},
\end{equation}
where we express the gas density as Hill density per dust-to-gas ratio $\varepsilon$. Combining both expressions results in:
\begin{equation}
	\lFree = \frac{\scaleheight \St}{3 f \varepsilon}
		\label{eq:lFreea}
\end{equation}
With our run parameters $\varepsilon = 3$ and $\St= 0.1$ this relates to 
\begin{equation}
	\lFree = \frac{0.01}{f} \scaleheight.
\end{equation}
This means that for all simulations, inside their initial homogeneous particle distribution the mean free path is larger than the smallest expected critical length of $l_c \simeq 0.004\scaleheight$. Within late stage particle overdensities with $f\geq10$ this now changes to $\lFree\geq0.001\scaleheight$, but the length scale of the overdensities is still smaller around $l \approx 0.0001\scaleheight$. 

We conclude that in all clumps found in our simulations the mean free path is equal to or larger than the clump size itself. When the mean free path indeed gets comparable to the clump size but the particle rms speed is less than the collapse velocity of the clump, then the collision timescale will still be longer than the collapse timescale. Our derivation here is equivalent to the discussion by \citet{2007Icar..192..588Y}.%

\subsection{Comparison to estimates in the literature}
\correctedold{
\citep{Nesvorny2010} find that for a KBO with a radius of 250 km at 30 $AU$ in a Minimum Mass Solar Nebula \citet{Hayashi1981} with 10 $g/cm^2$ local surface density the ratio between collision time and friction time should be
\begin{equation}
    \frac{\tau_\mathrm{coll}}{\tau_\mathrm{f}} \approx 0.05 \sqrt{\frac{R^2}{30 {\rm AU}}} \frac{250 {\rm km}}{R_{\rm eq}} f_H^{7/2}, 
\end{equation}
which would define the radius at which friction and collisions are equal to 12.5 km or 25 km in diameter. This is smaller than we would have estimated above, so we recapitulated their estimate. They used solid density of $2 g / cm^3$ and some additional order of magnitude short cuts.

In communication with the authors of \citet{Nesvorny2010} we found that for the nebula models in this paper the critical size to have collisions dominate over friction to be larger than 100 km (See Fig.\ \ref{fig:aeq2}) and smaller than the 500 km considered in \citet{Nesvorny2010}.}

It is therefore safe to neglect collisions in the present work (with $\varepsilon = 3$). In follow up three-dimensional studies we will treat them correctly in order to get a better understanding on the final outcome of planetesimals (e.g.\ multiplicity and spin rate) from the described process of self-gravity. 

\bibliographystyle{aasjournal}


\begin{thebibliography}{}
\expandafter\ifx\csname natexlab\endcsname\relax\def\natexlab#1{#1}\fi
\providecommand{\url}[1]{\href{#1}{#1}}
\providecommand{\dodoi}[1]{doi:~\href{http://doi.org/#1}{\nolinkurl{#1}}}
\providecommand{\doeprint}[1]{\href{http://ascl.net/#1}{\nolinkurl{http://ascl.net/#1}}}
\providecommand{\doarXiv}[1]{\href{https://arxiv.org/abs/#1}{\nolinkurl{https://arxiv.org/abs/#1}}}

\bibitem[{{Abod} {et~al.}(2018){Abod}, {Simon}, {Li}, {Armitage}, {Youdin}, \&
  {Kretke}}]{Abod2018}
{Abod}, C.~P., {Simon}, J.~B., {Li}, R., {et~al.} 2018, arXiv e-prints.
\newblock \doarXiv{1810.10018}

\bibitem[{{Andrews} {et~al.}(2010){Andrews}, {Wilner}, {Hughes}, {Qi}, \&
  {Dullemond}}]{Andrews2010}
{Andrews}, S.~M., {Wilner}, D.~J., {Hughes}, A.~M., {Qi}, C., \& {Dullemond},
  C.~P. 2010, ApJ, 723, 1241, \dodoi{10.1088/0004-637X/723/2/1241}

\bibitem[{{Asphaug} \& {Benz}(1996)}]{Asphaug1996}
{Asphaug}, E., \& {Benz}, W. 1996, \icarus, 121, 225,
  \dodoi{10.1006/icar.1996.0083}

\bibitem[{{Birnstiel} {et~al.}(2012){Birnstiel}, {Klahr}, \&
  {Ercolano}}]{Birnstiel2012}
{Birnstiel}, T., {Klahr}, H., \& {Ercolano}, B. 2012, \aap, 539, A148,
  \dodoi{10.1051/0004-6361/201118136}

\bibitem[{{Brandenburg}(2001)}]{Brandenburg2001}
{Brandenburg}, A. 2001, \apj, 550, 824, \dodoi{10.1086/319783}

\bibitem[{{Brandenburg} \& {Dobler}(2002)}]{Brandenburg2002}
{Brandenburg}, A., \& {Dobler}, W. 2002, Computer Physics Communicationss, 147,
  471

\bibitem[{{Brandenburg} \& {Dobler}(2005)}]{Brandenburg2005}
---. 2005, Astronomische Nachrichten, 326, 787

\bibitem[{{Brandenburg} {et~al.}(1995){Brandenburg}, {Nordlund}, {Stein}, \&
  {Torkelsson}}]{Brandenburg1995}
{Brandenburg}, A., {Nordlund}, A., {Stein}, R.~F., \& {Torkelsson}, U. 1995,
  \apj, 446, 741, \dodoi{10.1086/175831}

\bibitem[{{Carrera} {et~al.}(2017){Carrera}, {Gorti}, {Johansen}, \&
  {Davies}}]{Carrera2017}
{Carrera}, D., {Gorti}, U., {Johansen}, A., \& {Davies}, M.~B. 2017, \apj, 839,
  16, \dodoi{10.3847/1538-4357/aa6932}

\bibitem[{Chandrasekhar(1967)}]{Chandrasekhar1967}
Chandrasekhar, S. 1967, Communications on Pure and Applied {\ldots}, xx, 251

\bibitem[{{Connelly} {et~al.}(2012){Connelly}, {Bizzarro}, {Krot}, {Nordlund},
  {Wielandt}, \& {Ivanova}}]{2012Sci...338..651C}
{Connelly}, J.~N., {Bizzarro}, M., {Krot}, A.~N., {et~al.} 2012, Science, 338,
  651, \dodoi{10.1126/science.1226919}

\bibitem[{{Cuzzi} {et~al.}(2010){Cuzzi}, {Hogan}, \& {Bottke}}]{Cuzzi2010}
{Cuzzi}, J.~N., {Hogan}, R.~C., \& {Bottke}, W.~F. 2010, \icarus, 208, 518,
  \dodoi{10.1016/j.icarus.2010.03.005}

\bibitem[{{Cuzzi} {et~al.}(2008){Cuzzi}, {Hogan}, \& {Shariff}}]{Cuzzi2008}
{Cuzzi}, J.~N., {Hogan}, R.~C., \& {Shariff}, K. 2008, \apj, 687, 1432,
  \dodoi{10.1086/591239}

\bibitem[{{DeFelice} {et~al.}(2019){DeFelice}, {Friedrich}, {Ebel}, {Flores},
  \& {Weisberg}}]{2019LPI....50.2919D}
{DeFelice}, J.~D., {Friedrich}, J.~M., {Ebel}, D.~S., {Flores}, K.~E., \&
  {Weisberg}, M.~K. 2019, in Lunar and Planetary Science Conference, Lunar and
  Planetary Science Conference, 2919

\bibitem[{{Delbo{\textquoteright}} {et~al.}(2017){Delbo{\textquoteright}},
  {Walsh}, {Bolin}, {Avdellidou}, \& {Morbidelli}}]{Delbo2017}
{Delbo{\textquoteright}}, M., {Walsh}, K., {Bolin}, B., {Avdellidou}, C., \&
  {Morbidelli}, A. 2017, Science, 357, 1026, \dodoi{10.1126/science.aam6036}

\bibitem[{{Dittrich} {et~al.}(2013){Dittrich}, {Klahr}, \&
  {Johansen}}]{Dittrich2012}
{Dittrich}, K., {Klahr}, H., \& {Johansen}, A. 2013, \apj, 763, 117,
  \dodoi{10.1088/0004-637X/763/2/117}

\bibitem[{Dr{\k{a}}{\.z}kowska {et~al.}(2016)Dr{\k{a}}{\.z}kowska, Alibert, \&
  Moore}]{drazkowska2016}
Dr{\k{a}}{\.z}kowska, J., Alibert, Y., \& Moore, B. 2016, \aap, 594, A105

\bibitem[{Dubrulle {et~al.}(1995)Dubrulle, Morfill, \& Sterzik}]{Dubrulle1995}
Dubrulle, B., Morfill, G., \& Sterzik, M. 1995, Icarus, 114, 237

\bibitem[{{Gerbig} {et~al.}(2019){Gerbig}, {Lenz}, \& {Klahr}}]{gerbig2019}
{Gerbig}, K., {Lenz}, C.~T., \& {Klahr}, H. 2019, \aap, 629, A116,
  \dodoi{10.1051/0004-6361/201935278}

\bibitem[{{Gerbig} {et~al.}(2020){Gerbig}, {Murray-Clay}, {Klahr}, \&
  {Baehr}}]{Gerbig2020}
{Gerbig}, K., {Murray-Clay}, R.~A., {Klahr}, H., \& {Baehr}, H. 2020, \apj,
  895, 91, \dodoi{10.3847/1538-4357/ab8d37}

\bibitem[{{Goldreich} \& {Lynden-Bell}(1965)}]{Goldreich1965}
{Goldreich}, P., \& {Lynden-Bell}, D. 1965, \mnras, 130, 125,
  \dodoi{10.1093/mnras/130.2.125}

\bibitem[{{Goldreich} \& {Ward}(1973)}]{GoldreichWard1973}
{Goldreich}, P., \& {Ward}, W.~R. 1973, \apj, 183, 1051

\bibitem[{{Hartlep} \& {Cuzzi}(2020)}]{HartlepCuzzi2020}
{Hartlep}, T., \& {Cuzzi}, J.~N. 2020, arXiv e-prints, arXiv:2002.06321.
\newblock \doarXiv{2002.06321}

\bibitem[{{Hawley} \& {Balbus}(1992)}]{Hawley1992}
{Hawley}, J.~F., \& {Balbus}, S.~A. 1992, \apj, 400, 595,
  \dodoi{10.1086/172021}

\bibitem[{Hayashi(1981)}]{Hayashi1981}
Hayashi, C. 1981, Progress of Theoretical Physics Supplement, 70, 35,
  \dodoi{10.1143/PTPS.70.35}

\bibitem[{Jansson {et~al.}(2017)Jansson, Johansen, Syed, \&
  Blum}]{Wahlberg2017b}
Jansson, K.~W., Johansen, A., Syed, M.~B., \& Blum, J. 2017, The Astrophysical
  Journal, 835, 109, \dodoi{10.3847/1538-4357/835/1/109}

\bibitem[{{Jeans}(1902)}]{Jeans1902}
{Jeans}, J.~H. 1902, Philosophical Transactions of the Royal Society of London
  Series A, 199, 1, \dodoi{10.1098/rsta.1902.0012}

\bibitem[{{Johansen} {et~al.}(2014){Johansen}, {Blum}, {Tanaka}, {Ormel},
  {Bizzarro}, \& {Rickman}}]{Johansen2014}
{Johansen}, A., {Blum}, J., {Tanaka}, H., {et~al.} 2014, Protostars and Planets
  VI, 547, \dodoi{10.2458/azu\_uapress\_9780816531240-ch024}

\bibitem[{{Johansen} {et~al.}(2006){Johansen}, {Klahr}, \&
  {Henning}}]{JohansenKlahrHenning2006}
{Johansen}, A., {Klahr}, H., \& {Henning}, T. 2006, \apj, 636, 1121

\bibitem[{{Johansen} {et~al.}(2015){Johansen}, {Mac Low}, {Lacerda}, \&
  {Bizzarro}}]{Johansen2015}
{Johansen}, A., {Mac Low}, M.-M., {Lacerda}, P., \& {Bizzarro}, M. 2015,
  Science Advances, 1, 1500109, \dodoi{10.1126/sciadv.1500109}

\bibitem[{{Johansen} {et~al.}(2007{\natexlab{a}}){Johansen}, {Oishi}, {Mac
  Low}, {Klahr}, {Henning}, \& {Youdin}}]{2007Natur.448.1022J}
{Johansen}, A., {Oishi}, J.~S., {Mac Low}, M.-M., {et~al.} 2007{\natexlab{a}},
  \nat, 448, 1022, \dodoi{10.1038/nature06086}

\bibitem[{{Johansen} {et~al.}(2007{\natexlab{b}}){Johansen}, {Oishi}, {Mac
  Low}, {Klahr}, {Henning}, \& {Youdin}}]{2007arXiv0708.3893J}
---. 2007{\natexlab{b}}, arXiv e-prints, arXiv:0708.3893.
\newblock \doarXiv{0708.3893}

\bibitem[{{Johansen} \& {Youdin}(2007)}]{JohansenYoudin2007}
{Johansen}, A., \& {Youdin}, A. 2007, \apj, 662, 627

\bibitem[{{Klahr} \& {Bodenheimer}(2006)}]{2006ApJ...639..432K}
{Klahr}, H., \& {Bodenheimer}, P. 2006, \apj, 639, 432, \dodoi{10.1086/498928}

\bibitem[{{Klahr} {et~al.}(2018){Klahr}, {Pfeil}, \&
  {Schreiber}}]{2018haex.bookE.138K}
{Klahr}, H., {Pfeil}, T., \& {Schreiber}, A. 2018, {Instabilities and Flow
  Structures in Protoplanetary Disks: Setting the Stage for Planetesimal
  Formation} (Springer International Publishing AG, part of Springer Nature),
  138

\bibitem[{{Klahr} \& {Schreiber}(AAS25612)}]{KlahrSchreiber2020b}
{Klahr}, H., \& {Schreiber}, A. AAS25612, Submitted to ApJ

\bibitem[{{Kobayashi} {et~al.}(2016){Kobayashi}, {Tanaka}, \&
  {Okuzumi}}]{Kobayashi2016}
{Kobayashi}, H., {Tanaka}, H., \& {Okuzumi}, S. 2016, \apj, 817, 105,
  \dodoi{10.3847/0004-637X/817/2/105}

\bibitem[{{Lenz} {et~al.}(2019){Lenz}, {Klahr}, \& {Birnstiel}}]{Lenz_2019}
{Lenz}, C.~T., {Klahr}, H., \& {Birnstiel}, T. 2019, \apj, 874, 36,
  \dodoi{10.3847/1538-4357/ab05d9}

\bibitem[{{Lenz} {et~al.}(2020){Lenz}, {Klahr}, {Birnstiel}, {Kretke}, \&
  {Stammler}}]{Lenz2020}
{Lenz}, C.~T., {Klahr}, H., {Birnstiel}, T., {Kretke}, K., \& {Stammler}, S.
  2020, arXiv e-prints, arXiv:2006.08799.
\newblock \doarXiv{2006.08799}

\bibitem[{Levison {et~al.}(2011)Levison, Morbidelli, Tsiganis, Nesvorn{\`y}, \&
  Gomes}]{Levison2011}
Levison, H.~F., Morbidelli, A., Tsiganis, K., Nesvorn{\`y}, D., \& Gomes, R.
  2011, The Astronomical Journal, 142, 152

\bibitem[{{Lyra} {et~al.}(2008){Lyra}, {Johansen}, {Klahr}, \&
  {Piskunov}}]{Lyra2008}
{Lyra}, W., {Johansen}, A., {Klahr}, H., \& {Piskunov}, N. 2008, \aap, 479,
  883, \dodoi{10.1051/0004-6361:20077948}

\bibitem[{{Lyra} {et~al.}(2009){Lyra}, {Johansen}, {Zsom}, {Klahr}, \&
  {Piskunov}}]{Lyra2009}
{Lyra}, W., {Johansen}, A., {Zsom}, A., {Klahr}, H., \& {Piskunov}, N. 2009,
  \aap, 497, 869, \dodoi{10.1051/0004-6361/200811265}

\bibitem[{{Morbidelli} {et~al.}(2009){Morbidelli}, {Bottke}, {Nesvorn{\'y}}, \&
  {Levison}}]{Morbidelli2009}
{Morbidelli}, A., {Bottke}, W.~F., {Nesvorn{\'y}}, D., \& {Levison}, H.~F.
  2009, \icarus, 204, 558, \dodoi{10.1016/j.icarus.2009.07.011}

\bibitem[{Morbidelli {et~al.}(2007)Morbidelli, Tsiganis, Crida, Levison, \&
  Gomes}]{Morbidelli2007}
Morbidelli, A., Tsiganis, K., Crida, A., Levison, H.~F., \& Gomes, R. 2007, The
  Astronomical Journal, 134, 1790

\bibitem[{{Nakagawa} {et~al.}(1986){Nakagawa}, {Sekiya}, \&
  {Hayashi}}]{Nakagawa1986}
{Nakagawa}, Y., {Sekiya}, M., \& {Hayashi}, C. 1986, \icarus, 67, 375,
  \dodoi{10.1016/0019-1035(86)90121-1}

\bibitem[{{Nesvorn{\'y}} {et~al.}(2019){Nesvorn{\'y}}, {Li}, {Youdin}, {Simon},
  \& {Grundy}}]{2019NatAs.tmp..415N}
{Nesvorn{\'y}}, D., {Li}, R., {Youdin}, A.~N., {Simon}, J.~B., \& {Grundy},
  W.~M. 2019, Nature Astronomy, 415, \dodoi{10.1038/s41550-019-0806-z}

\bibitem[{{Nesvorn{\'y}} {et~al.}(2010){Nesvorn{\'y}}, {Youdin}, \&
  {Richardson}}]{Nesvorny2010}
{Nesvorn{\'y}}, D., {Youdin}, A.~N., \& {Richardson}, D.~C. 2010, \aj, 140,
  785, \dodoi{10.1088/0004-6256/140/3/785}

\bibitem[{{Offner} {et~al.}(2014){Offner}, {Clark}, {Hennebelle}, {Bastian},
  {Bate}, {Hopkins}, {Moraux}, \& {Whitworth}}]{2014prpl.conf...53O}
{Offner}, S.~S.~R., {Clark}, P.~C., {Hennebelle}, P., {et~al.} 2014, in
  Protostars and Planets VI, ed. H.~{Beuther}, R.~S. {Klessen}, C.~P.
  {Dullemond}, \& T.~{Henning}, 53

\bibitem[{{Ormel} \& {Klahr}(2010)}]{OrmelKlahr2010}
{Ormel}, C.~W., \& {Klahr}, H.~H. 2010, \aap, 520, A43,
  \dodoi{10.1051/0004-6361/201014903}

\bibitem[{{Pfeil} \& {Klahr}(2019)}]{Pfeil2019}
{Pfeil}, T., \& {Klahr}, H. 2019, \apj, 871, 150,
  \dodoi{10.3847/1538-4357/aaf962}

\bibitem[{{Raettig} {et~al.}(2015){Raettig}, {Klahr}, \& {Lyra}}]{Raettig2014}
{Raettig}, N., {Klahr}, H., \& {Lyra}, W. 2015, \apj, 804, 35,
  \dodoi{10.1088/0004-637X/804/1/35}

\bibitem[{Raymond \& Izidoro(2017)}]{raymond2017}
Raymond, S.~N., \& Izidoro, A. 2017, Science advances, 3, e1701138

\bibitem[{{Safronov}(1969)}]{Safronov1969}
{Safronov}, V.~S. 1969, {Evoliutsiia Doplanetnogo Oblaka.} (English transl.:
  Evolution of the protoplanetary cloud and formation of Earth and the planets,
  NASA Tech. Transl. F-677, Jerusalem: Israel Sci. Transl. 1972)

\bibitem[{{Sch{\"a}fer} {et~al.}(2017){Sch{\"a}fer}, {Yang}, \&
  {Johansen}}]{Schaefer2017}
{Sch{\"a}fer}, U., {Yang}, C.-C., \& {Johansen}, A. 2017, \aap, 597, A69,
  \dodoi{10.1051/0004-6361/201629561}

\bibitem[{Schoonenberg {et~al.}(2018)Schoonenberg, Ormel, \&
  Krijt}]{schoonenberg2018}
Schoonenberg, D., Ormel, C.~W., \& Krijt, S. 2018, Astronomy \& Astrophysics,
  620, A134

\bibitem[{{Schreiber} \& {Klahr}(2018)}]{Schreiber2018}
{Schreiber}, A., \& {Klahr}, H. 2018, \apj, 861, 47,
  \dodoi{10.3847/1538-4357/aac3d4}

\bibitem[{{Sekiya}(1983)}]{Sekiya1983}
{Sekiya}, M. 1983, Progress of Theoretical Physics, 69, 1116,
  \dodoi{10.1143/PTP.69.1116}

\bibitem[{{Shariff} \& {Cuzzi}(2015)}]{Shariff2015}
{Shariff}, K., \& {Cuzzi}, J.~N. 2015, \apj, 805, 42,
  \dodoi{10.1088/0004-637X/805/1/42}

\bibitem[{{Shi} \& {Chiang}(2013)}]{Shi2013}
{Shi}, J.-M., \& {Chiang}, E. 2013, \apj, 764, 20,
  \dodoi{10.1088/0004-637X/764/1/20}

\bibitem[{{Simon} {et~al.}(2016){Simon}, {Armitage}, {Li}, \&
  {Youdin}}]{Simon2016}
{Simon}, J.~B., {Armitage}, P.~J., {Li}, R., \& {Youdin}, A.~N. 2016, \apj,
  822, 55, \dodoi{10.3847/0004-637X/822/1/55}

\bibitem[{{Simon} {et~al.}(2017){Simon}, {Armitage}, {Youdin}, \&
  {Li}}]{Simon2017}
{Simon}, J.~B., {Armitage}, P.~J., {Youdin}, A.~N., \& {Li}, R. 2017, \apjl,
  847, L12, \dodoi{10.3847/2041-8213/aa8c79}

\bibitem[{{Singer} {et~al.}(2019){Singer}, {McKinnon}, {Gladman},
  {Greenstreet}, {Bierhaus}, {Stern}, {Parker}, {Robbins}, {Schenk}, {Grundy},
  {Bray}, {Beyer}, {Binzel}, {Weaver}, {Young}, {Spencer}, {Kavelaars},
  {Moore}, {Zangari}, {Olkin}, {Lauer}, {Lisse}, {Ennico}, {New Horizons
  Geology}, {New Horizons Surface Composition Science Theme Team}, \& {New
  Horizons Ralph and LORRI Teams}}]{Singer2019}
{Singer}, K.~N., {McKinnon}, W.~B., {Gladman}, B., {et~al.} 2019, Science, 363,
  955, \dodoi{10.1126/science.aap8628}

\bibitem[{{Squire} \& {Hopkins}(2018)}]{2018MNRAS.477.5011S}
{Squire}, J., \& {Hopkins}, P.~F. 2018, \mnras, 477, 5011,
  \dodoi{10.1093/mnras/sty854}

\bibitem[{Stephan \& Docter(2015)}]{Stephan:202326}
Stephan, M., \& Docter, J. 2015, Journal of large-scale research facilities, 1,
  A1, \dodoi{10.17815/jlsrf-1-18}

\bibitem[{{Stern} {et~al.}(2019){Stern}, {Spencer}, {Weaver}, {Olkin}, {Moore},
  {Grundy}, {Gladstone}, {McKinnon}, {Cruikshank}, {Young}, {Elliott},
  {Verbiscer}, {Parker}, \& {New Horizons Team}}]{Stern2019}
{Stern}, S.~A., {Spencer}, J.~R., {Weaver}, H.~A., {et~al.} 2019, in Lunar and
  Planetary Science Conference, Lunar and Planetary Science Conference, 1742

\bibitem[{{Tsirvoulis} {et~al.}(2018){Tsirvoulis}, {Morbidelli}, {Delbo}, \&
  {Tsiganis}}]{Tsirvoulis2016}
{Tsirvoulis}, G., {Morbidelli}, A., {Delbo}, M., \& {Tsiganis}, K. 2018,
  \icarus, 304, 14, \dodoi{10.1016/j.icarus.2017.05.026}

\bibitem[{{Wahlberg Jansson} \& {Johansen}(2017)}]{Wahlberg2017}
{Wahlberg Jansson}, K., \& {Johansen}, A. 2017, \mnras, 469, S149,
  \dodoi{10.1093/mnras/stx1470}

\bibitem[{Walsh {et~al.}(2011)Walsh, Morbidelli, Raymond, O'Brien, \&
  Mandell}]{Walsh2011}
Walsh, K.~J., Morbidelli, A., Raymond, S.~N., O'Brien, D.~P., \& Mandell, A.~M.
  2011, Nature, 475, 206

\bibitem[{{Ward}(2000)}]{Ward2000}
{Ward}, W.~R. 2000, {On Planetesimal Formation: The Role of Collective Particle
  Behavior}, ed. R.~M. {Canup}, K.~{Righter}, \& {et al.}, 75--84

\bibitem[{{Youdin} \& {Johansen}(2007)}]{YoudinJohansen2007}
{Youdin}, A., \& {Johansen}, A. 2007, \apj, 662, 613

\bibitem[{{Youdin} \& {Goodman}(2005)}]{YoudinGoodman2005}
{Youdin}, A.~N., \& {Goodman}, J. 2005, \apj, 620, 459

\bibitem[{{Youdin} \& {Lithwick}(2007)}]{2007Icar..192..588Y}
{Youdin}, A.~N., \& {Lithwick}, Y. 2007, \icarus, 192, 588,
  \dodoi{10.1016/j.icarus.2007.07.012}

\end{thebibliography}



\end{document}